\documentclass[aps,pra,twocolumn,superscriptaddress]{revtex4-1}
\pdfoutput=1
\usepackage{graphicx}
\usepackage{amssymb}
\usepackage{amsmath}
\usepackage{color}
\usepackage[unicode=true,
 bookmarks=false,
 breaklinks=false,pdfborder={0 0 1},backref=false,colorlinks=true]
 {hyperref}
\hypersetup{
 linkcolor=blue,citecolor=blue,urlcolor=blue,linkcolor=blue,citecolor=blue,urlcolor=blue}

\begin{document}

\title{A brief review of topological photonics in one, two, and three dimensions}

\author{Zhihao Lan}
\affiliation{Department of Electronic and Electrical Engineering, University College London, Torrington Place, London, WC1E 7JE, United Kingdom}
\author{Menglin L. N. Chen}
\affiliation{Department of Physics, The University of Hong Kong, Hong Kong, China}
\author{Fei Gao}
\affiliation{Interdisciplinary Center for Quantum Information, State Key Laboratory of Modern Optical Instrumentation, ZJU-Hangzhou Global Science and Technology Innovation Center, College of Information Science and Electronic Engineering, Zhejiang University, Hangzhou, China}
\author {Shuang Zhang}
\affiliation{Department of Physics, The University of Hong Kong, Hong Kong, China}
\affiliation{Department of Electrical and Electronic Engineering, The University of Hong Kong, Hong Kong, China}
\author{Wei E. I. Sha}
\email{weisha@zju.edu.cn}
\affiliation{Key Laboratory of Micro-nano Electronic Devices and Smart Systems of Zhejiang Province, College of Information Science and Electronic Engineering, Zhejiang University, Hangzhou 310027, China}

\date{\today}

\begin{abstract}
Topological photonics has attracted increasing attention in recent years due to the unique opportunities it provides to manipulate light in a robust way immune to disorder and defects.  Up to now, diverse photonic platforms, rich physical mechanisms and fruitful device applications have been proposed for topological photonics, including one-way waveguide, topological lasing, topological nanocavity, Dirac and Weyl points, Fermi arcs, nodal lines, etc. In this review, we provide an introduction to the field of topological photonics through the lens of topological invariants and bulk-boundary correspondence in one, two, and three dimensions, which may not only offer a unified understanding about the underlying robustness of diverse and distinct topological phenomena of light, but could also inspire further developments by introducing new topological invariants and unconventional bulk-boundary correspondence to the research of topological photonics.
\end{abstract}

\maketitle

\section{Introduction}
The field of topological photonics explores topology-related phenomena within the theoretical framework of Maxwell's equations
for electromagnetic waves and is largely motivated by the earlier developments in condensed matter physics. However, due to the fundamental difference between photons and electrons, i.e., while electrons are fermions obeying the Fermi-Dirac statistics, photons are bosons governed by the Bose-Einstein statistics, new physics and problems specific to the bosonic nature of light not only motivate photonics researchers to find {\it new ways to do old things}, but also revolutionize our fundamental understanding of how light could be manipulated in a completely new way. Through the bulk-edge correspondence principle, topological bulk properties of photonic systems could lead to nontrivial behaviors of light along the edge of the systems, which usually are robust against certain types of defects and disorders. This prominent feature is very promising for a broad range of photonic applications in the near future. The field of topological photonics started in 2008 from a work by Haldane and Raghu \cite{Haldane08PRL}, who proposed to mimic the quantum Hall effect of two-dimensional (2D) electron gases in magnetic fields using light. The idea was experimentally realized soon afterwards \cite{Wang09nature} in a gyromagnetic photonic crystal composed of ferrite rods in the microwave regime. After this fundamental breakthrough, great developments for achieving topological states of electromagnetic waves have been made over the past 14 years, which were motivated by both the topological states in condensed matter physics and the specific features of electromagnetic waves. For example, while the earlier works \cite{Haldane08PRL,Wang09nature} have explored magnetic-optical effects of gyromagnetic materials to mimic quantum Hall states, magnetic-optical effects in general are very weak at optical frequencies, and this motivated the ideas of creating artificial magnetic fields for photons \cite{Fang12NatPho_magnetic,Hafezi11NatPhy_resonator}.  Several recent advances in photonic topological insulators, such as $Z_2$ \cite{Khanikaev13NatMat}, Floquet \cite{Rechtsman13Nature}, valley-Hall \cite{MaShvets16NJP}, and second-order \cite{Xie18PRB_corner}, were largely motivated by the relevant concepts developed in condensed matter physics. Especially, the richness of topological physics in three-dimensional (3D) condensed matter systems has provided many exciting new opportunities for 3D topological photonics \cite{Kim20Light_review,Xie21Frontier_3Drev,Park22NanoPhot_3Drev}.

%%%%%%

The field of topological photonics has been constantly reviewed in the past due to its fast developments \cite{Lu14NatPho_review,Khanikaev17NatPho_review,Xie18OE_review,Ozawa19RMP_review,Tang22LPR_review}. The bulk-edge correspondence is an important principle to understand topological phenomena, which states that if the bulk of a system hosts a nontrivial topological invariant, then this invariant would manifest itself in the edge of the system. This principle not only reveals a deep connection between physics in different dimensions, but also provides a unique perspective to understand the diverse phenomena in topological photonic systems in a unified way. Note that proving this principle in a mathematically rigorous way in general is hard considering the diverse photonic systems available \cite{Silveirinha19PRX_bulk-edge} and in certain cases, one could have anomalous bulk-edge correspondence \cite{Silveirinha16PRB_bulk-edge,Tauber20PRR_bulk-edge} or even the violations of the bulk-edge correspondence \cite{Gangaraj20PRL_bulk-edge}. In this review, we present a brief introduction to the field of topological photonics, covering most of the recent developments to date in one, two, and three dimensions. We would like to note that due to the fundamental difference between electrons and photons, in finite systems with boundaries, the conditions for the existence of edge states of the two systems are different. As we know, electrons can not escape the boundary of a system due to the work function, which provides a confining potential and as such, the wave functions of electrons are evanescent beyond the system boundary, i.e., air is naturally a topologically trivial medium for electrons. In contrast, light can propagate and exist in free space, i.e., in general, no confining potential exists between photonic media and free space. Different methods to confine light inside photonic media exist, e.g., using perfect electric conductors (or metals), the bandgap of a photonic crystal or exploiting the light cone effect. This difference between electronic and photonic systems can make the methods to form topological edge states in photonic systems more involved, but which will also provide new opportunities allowing the further control of these photonic topological edge states.

This review is organized as follows: In Sec. \ref{sec2}, we discuss the Maxwell-Schr\"odinger correspondence and some symmetry properties of electromagnetic waves in media. In Sec. \ref{sec3}, we present the prototypical model for topological physics in 1D, i.e., the Su-Schrieffer-Heeger (SSH) model, which has been used extensively in photonics for various topological applications. In Sec. \ref{sec4}, we overview four main categories of photonic topological states in 2D, i.e., quantum Hall states, quantum spin Hall states, quantum valley Hall states and second-order topological corner states. In Sec. \ref{sec5}, we discuss various topological states in 3D photonic systems, which in general could be classified as gapped or gapless. For gapless topological states, depending on the dimensions of the band crossing, we describe two important categories of nodal physics, i.e., nodal points and nodal lines. We summarize and give some future perspectives in Sec. \ref{sec6}.

%%%%%%%%%%%%%%%%%%%%%%%%%%%%%%%%%%%%%%%%
\section{Maxwell-Schr\"odinger correspondence and symmetries of electromagnetism} \label{sec2}
%%%%%%%%%%%%%%%%%%%%%%%%%%%%%%%%%%%%%%%%

Maxwell's equations could be formulated in the Schr\"odinger form \cite{Lu14NatPho_review,NittisLein19ATMP}, such that the concepts and toolboxes developed for topological electronic materials could be carried over to topological photonic systems. Maxwell's equations in a medium (without free charges and currents) are described by

\begin{gather}
\nabla \cdot \mathbf{D} =0, \hspace{0.5cm} \nabla \cdot \mathbf{B} =0 \\
\nabla \times \mathbf{E} =-\frac{\partial \mathbf{B}}{\partial t}, \hspace{0.5cm} \nabla \times \mathbf{H} =\frac{\partial \mathbf{D}}{\partial t}
 \end{gather}
which could be recast in the following  Schr\"odinger form
\begin{gather}
i\frac{\partial}{\partial t} \Psi =\hat{M} \Phi
\end{gather}
with $\Psi = \hat{N} \Phi$ and $(\Psi, \Phi, \hat{M}, \hat{N} )$ defined by
\begin{gather}
\Psi\equiv (\mathbf{D}, \mathbf{B})^T, \hspace{0.5cm} \Phi\equiv (\mathbf{E}, \mathbf{H})^T\\
\hat{M} =
\begin{pmatrix}
0 & i \nabla \times  \\
-i\nabla \times& 0
\end{pmatrix}, \hspace{0.5cm}
 \hat{N} =
\begin{pmatrix}
\epsilon & \xi  \\
\zeta & \mu
\end{pmatrix}
 \end{gather}
 where, $\epsilon$ and $\mu$ are the electric permittivity and magnetic permeability tensors, respectively, whereas $\xi$ and $\zeta$ are the bianisotropic magnetoelectric coupling tensors.
 The time-reversal ($T$), inversion ($P$) and duality ($S$) transformations of Maxwell's equations are given by
 \begin{gather}
 T: t \rightarrow -t, \hspace{0.2cm}(\mathbf{E},\mathbf{H})  \rightarrow (\mathbf{E},-\mathbf{H}) \\
 P: (x,y,z)\rightarrow -(x,y,z), \hspace{0.2cm}(\mathbf{E},\mathbf{H})  \rightarrow (-\mathbf{E},\mathbf{H}) \\
 S: (\mathbf{E}, \mathbf{H}) \rightarrow (\mathbf{H}, -\mathbf{E})
  \end{gather}
One can show that the above transformations are symmetries of  Maxwell's equations in vacuum, i.e., $\mathbf{D}=\epsilon_0\mathbf{E}$,
$\mathbf{B}=\mu_0\mathbf{H}$ and $\xi=\zeta=0$, where $\epsilon_0$ and $\mu_0$ are the vacuum permittivity and permeability, respectively. However, in a medium, the nontrivial material response, i.e., the explicit form of  $\hat{N}$  could break these symmetries. In the following, we briefly discuss the time-reversal transformation as it plays an important role in various photonic quantum Hall related states. First, consider a gyrotropic medium, i.e., gyroelectric:
\begin{gather}
 \epsilon = \begin{pmatrix}
 \epsilon_{xx} & i\epsilon_{b} &0  \\
-i\epsilon_{b} & \epsilon_{yy} &0  \\
0 & 0 & \epsilon_{zz}
\end{pmatrix}, \hspace{0.5cm} \mu=\mu_0
\label{ge}
\end{gather}
or gyromagnetic:
\begin{gather}
 \epsilon=\epsilon' (\text{scalar}), \hspace{0.5cm}\mu = \begin{pmatrix}
 \mu_{xx} & i\mu_{b} &0  \\
-i\mu_{b} & \mu_{yy} &0  \\
0 & 0 & \mu_{zz}
\end{pmatrix}
\label{gu}
\end{gather}
with $\xi=\zeta=0$, which could be realized by applying an external static magnetic field along $z$ of the gyrotropic material.  Since time-reversal is an antilinear transformation composed of a linear operator and the complex conjugation ($^*$), it changes the imaginary unit $i$ to $-i$. As such, the existence of the off-diagonal terms, $ i\epsilon_{b}$ or $ i\mu_{b}$ in Eq. (\ref{ge}) or (\ref{gu}) breaks the time-reversal symmetry of the system. In other words, one would need $\epsilon=\epsilon^*$ and $\mu=\mu^*$ in order for time-reversal to be preserved. In view of this, gyrotropic materials have been routinely used to realize one-way propagation of electromagnetic waves by mimicking the quantum Hall effect.

Furthermore, apart from breaking the time-reversal symmetry, preserving it can also lead to new topological phases. Comparing the time-reversal operators for electrons ($\mathcal{T}_{\textrm{e}}$) and photons ($\mathcal{T}_{\textrm{p}}$),
\begin{gather}
\mathcal{T}_{\textrm{e}}=i\sigma_y K \\
\mathcal{T}_{\textrm{p}}=\tau_z K
\end{gather}
where $K$ denotes the complex conjugation operator, $\sigma_y$ and $\tau_z$ are Pauli matrices acting on the spin ($\{\uparrow, \downarrow\}$) of electrons and the electromagnetic components ($ \{\mathbf{E}, \mathbf{H} \}$) of photons, respectively.  One can explicitly check that $\mathcal{T}_{\textrm{e}}\mathcal{T}_{\textrm{e}}=-1$ and $\mathcal{T}_{\textrm{p}}\mathcal{T}_{\textrm{p}}=+1$. Thus unlike electrons, where the Kramers degeneracy due to time-reversal symmetry could lead to $\mathcal{Z}_2$ topological phase (also known as quantum spin Hall phase in 2D), the time-reversal symmetry of photons alone is not sufficient to protect such phase and additional symmetry, such as crystalline symmetry, is needed to construct a photonic quantum spin Hall phase.

%%%%%%%%%%%%%%%%%%%%%%%%%
\section{Topological photonics in 1D}\label{sec3}
%%%%%%%%%%%%%%%%%%%%%%%%%%

\begin{figure}
\includegraphics[width=\columnwidth]{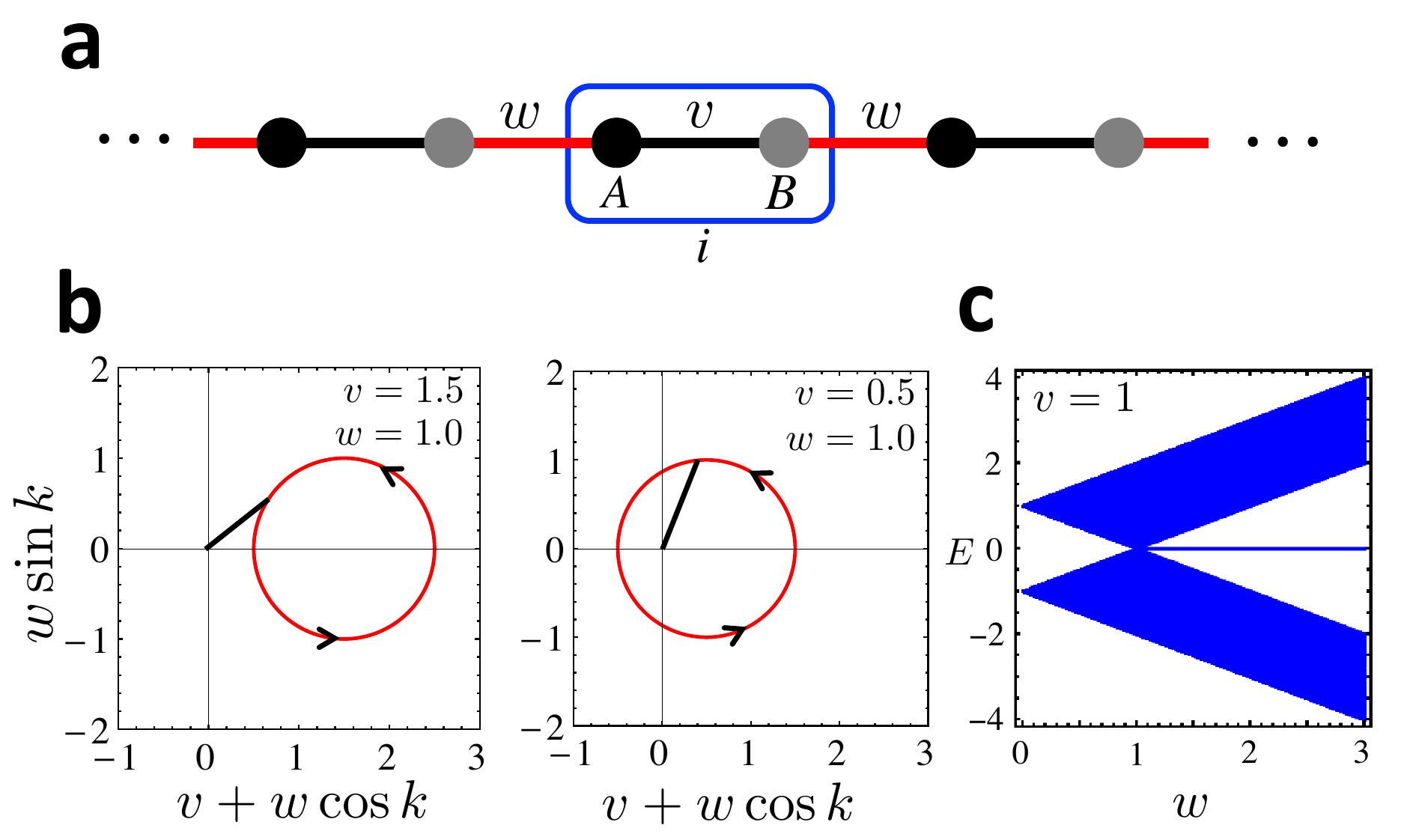}
\caption{ 1D SSH model and its topological properties. (a) The unit cell $i$ of the 1D SSH model contains two lattice sites A and B, and the model has two hopping amplitudes, the intra-cell hopping $v$ and the inter-cell hopping $w$. (b) Winding behaviors of the trajectory $(v+w\cos k,w\sin k)$ around the origin $(0,0)$ as $k$ changes from $-\pi$ to $+\pi$ when $v>w$ (left) and $v<w$ (right). (c) Energy spectrum of a finite 1D SSH chain (with 100 unit cells) as a function of $w$ at $v=1$. When $w>v$, zero-energy edge states appear within the bandgap.}
\label{figs:fig1}
\end{figure}

\begin{figure*}
\includegraphics[width=0.9\textwidth]{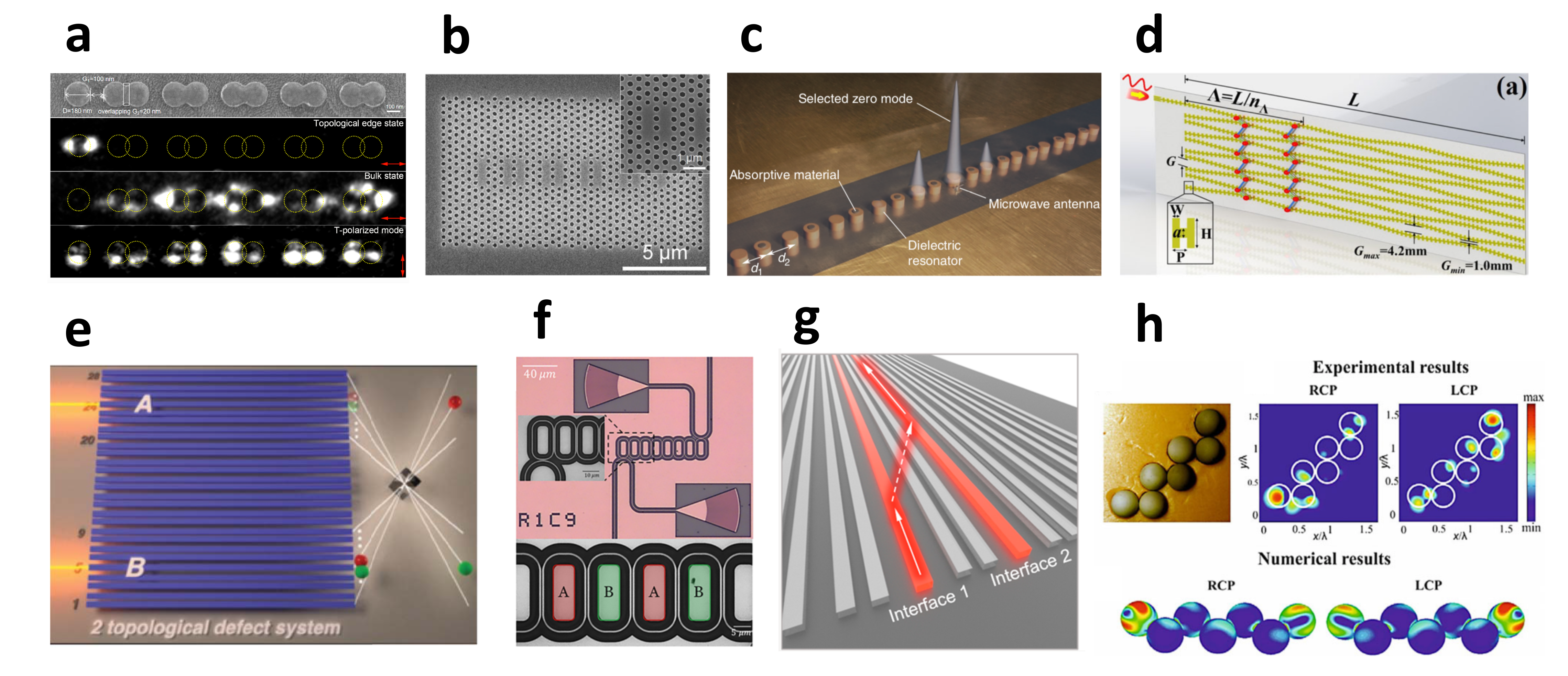}
\caption{ 1D photonic SSH systems and their applications. (a) Near-field imaging of topological edge states in plasmonic nanochains. Reproduced with permission from~\cite{Yan21NanoLett_time}, Copyright (2021), under CC BY-NC-ND. (b) Lasing at the topological edge states in a photonic crystal nanocavity dimer array. Reproduced with permission from~\cite{Han19Light_lasing}, Copyright (2019), under CC BY 4.0. (c) Selective enhancement of interface states in a dielectric resonator chain. Reproduced with permission from~\cite{Poli15NC_nonH}, Copyright (2015), under CC BY 4.0. (d) Periodically bended ultrathin metallic waveguides for the observation of anomalous $\pi$ modes via photonic floquet engineering. Reproduced with permission from~\cite{Cheng19PRL_drivenSSH}, Copyright (2019) by the American Physical Society. (e) Protection of biphoton entanglement in an array of silicon nanowires. Reproduced with permission from~\cite{Redondo18Science_biphoton}, Copyright (2018) by the American Association for the Advancement of Science. (f) Photonic topological baths for quantum simulation. Reproduced with permission from~\cite{Saxena22ACSpho_bath}, Copyright (2022) by the American Chemical Society. (g) Two coupled topological interfaces in waveguide arrays. Reproduced with permission from~\cite{Wang20NanoLett_interaction}, Copyright (2020) by the American Chemical Society. (h) Selective excitation of topological edge states controlled by the handedness of incident light. Reproduced with permission from~\cite{Slobozhanyuk16LPR}, Copyright (2016) by John Wiley and Sons. }
\label{figs:fig2}
\end{figure*}

The prototypical model for topological physics in 1D is the SSH model \cite{SSH79PRL}, which describes the hopping of a particle in a 1D dimerized lattice with alternating strong and weak hopping amplitudes (see Fig.\ref{figs:fig1}a). The unit cell of the dimerized lattice contains two sites $A$ and $B$, and the Hamiltonian describing the system is given by
\begin{gather}
\hat{H} = \sum_i v(\hat{c}_{i,A}^{\dagger} \hat{c}_{i,B} +\hat{c}_{i,B}^{\dagger}\hat{c}_{i,A}) \nonumber  \\
+w(\hat{c}_{i+1,A}^{\dagger} \hat{c}_{i,B} +\hat{c}_{i-1,B}^{\dagger} \hat{c}_{i,A})
\end{gather}
where $\hat{c}_{i,A/B}^{\dagger}$ ($\hat{c}_{i,A/B}$) is the creation (annihilation) operator of a particle on the site $A/B$ within the unit cell $i$, whereas $v$ and $w$ are the intra-cell and inter-cell hopping amplitudes, respectively. Applying the Fourier transform $\hat{c}_{R,A/B}^{\dagger}=\frac{1}{\sqrt{N}}\sum_k e^{-ikR}\hat{c}_{k,A/B}^{\dagger}$, where $N$ is the number of lattice sites, and using the identity $\frac{1}{N}\sum_R e^{ikR}=\delta_{k0}$, one can write the Hamiltonian in momentum space  as $\hat{H} = \hat{C}_{k}^{\dagger} H(k)  \hat{C}_{k}$ with $ \hat{C}_{k} = (\hat{c}_{k,A}, \hat{c}_{k,B})^T$ and
\begin{gather}
H(k)=\begin{pmatrix}
0 & v+we^{-ik} \\
v+we^{ik} & 0 \end{pmatrix}
\end{gather}
To make the physics more transparent, we could introduce a complex number $h(k)\equiv h_x(k)+ih_y(k)=|h(k)|e^{i\phi(k)}$ with $h_x(k)=v+w\cos k$ and $h_y(k)=w\sin k$. Then $H(k)$ reduces to
\begin{gather}
H(k)=|h(k)| \begin{pmatrix}
0 & e^{-i\phi(k)} \\
e^{i\phi(k)}& 0 \end{pmatrix}
\end{gather}

Now it is straightforward to show that the eigenvalues and eigenfunctions of $H(k)$ are
\begin{gather}
E_{\pm}(k)=\pm |h(k)| = \pm \sqrt{v^2+2vw\cos(k)+w^2} \\
\Psi_{\pm}(k)=\frac{1}{\sqrt{2}}\begin{pmatrix} \pm e^{-i\phi(k)} \\ 1\end{pmatrix}
\end{gather}
where the energy spectrum is gapped unless $v=w$. The Zak phase \cite{Zak89PRL} could be calculated as
\begin{gather}
\gamma=i\int_{-\pi}^{+\pi} \langle \Psi_{\pm}(k)|\partial_k |  \Psi_{\pm}(k)\rangle dk=\frac{1}{2}\int_{-\pi}^{+\pi} \frac{\partial \phi(k)}{\partial k} dk \nonumber \\ =\frac{1}{2} [\phi(+\pi)-\phi(-\pi)]
\end{gather}
whose physics now becomes apparent, i.e.,  if the trajectory of $h(k)$ encloses the origin when $k$ changes from $-\pi$ to $\pi$,  $\gamma =\pi$; otherwise $\gamma = 0$. From $h_x(k)=v+w\cos k$ and $h_y(k)=w\sin k$, it is easy to see from Fig.\ref{figs:fig1}b that when $v>w$, the close loop from $h(k)$ as $k$ changes does not enclose the origin and as such $\gamma=0$, whereas when $v<w$, the loop does enclose the origin and consequently, $\gamma=\pi$. At $v=w$, the bandgap closes, signaling a topological phase transition.

The SSH Hamiltonian $H(k)$ has trivial time-reversal symmetry implemented by $\mathcal{T} =K$, i.e., $\mathcal{T} H(k) \mathcal{T}^{-1}=H(-k)$, inversion symmetry implemented by $\mathcal{I} =\sigma_x$, i.e., $\mathcal{I} H(k) \mathcal{I}^{-1}=H(-k)$ and chiral symmetry implemented by $\mathcal{C} =\sigma_z$, i.e., $\mathcal{C} H(k) \mathcal{C}^{-1}=-H(k)$. The chiral symmetry will pin the topological edge states that emerge in a finite lattice when $w>v$ at zero energy (see Fig.\ref{figs:fig1}c).

Despite the simplicity of the 1D SSH model, the system platforms and physics that could be explored with it are very rich. For example, the 1D SSH model has been studied in diverse photonic platforms, such as, photonic superlattice in a photorefractive material \cite{Malkova09OL}, photonic crystal of alternating dielectric slabs \cite{Henriques20PRA_Tamm} or nanobeam cavities \cite{Gong21SciRep_nanobeam}, quantum emitters interacting with a waveguide \cite{Bello19SciAdv}, split-ring-resonators \cite{Jiang18OE_split},
dielectric nanoparticles \cite{Slobozhanyuk15PRL,Kruk17small}, array of coupled waveguides \cite{Redondo16PRL,Naz18PRA_stretch,Cheng18OE_Hbar,Chen19AP_tamm,Savelev20PRB_2mode,Song20LPR_waveguide,Jiao21PRL_inversionSSH}, plasmonic systems \cite{Poddubny14ACSpho_zigzag,Ling15OE,Cheng15LPR,Sinev15Nanoscale,Pocock18ACSpho_retard,Rappoport21ACSpho_graphene,Zhang21JAP}, and exciton-polaritons \cite{Solnyshkov16PRL_Zurek,Whittaker19PRB_SOC,Pickup20NC_nonH,Su21SciAdv,Pieczarka21OPtica_polariton}. The topological edge states of 1D photonic SSH systems have also been observed through imaging, such as, spectral imaging \cite{Bleckmann17PRB_imaging},  near-field imaging  \cite{Yan21NanoLett_time}(Fig.\ref{figs:fig2}a) and  far-field optical imaging \cite{Moritake2022NanoPhot_imaging}.

Moreover, various applications based on the 1D SSH model have been proposed in the past, such as, lasing \cite{Jean17NatPho_lasing,Zhao18NC_lasing,Parto18PRL_lasing,Ota18ComPhy_lasing,Han19Light_lasing,Gagel22ACSpho_switch}, non-Hermitian physics \cite{Schomerus13OL_nonH,Poli15NC_nonH,Zeuner15PRL_nonH,Pan18NC,Song19PRL_recover,Zhu20PRR_skin,Lin21OE_squareroot}, Floquet dynamics \cite{Cheng19PRL_drivenSSH,Petracek20PRA_drivenSSH},
quantum information of biphoton states \cite{Redondo18Science_biphoton,Klauck21PhoRes_biphoton}, quantum photonic baths \cite{Saxena22ACSpho_bath}, nonreciprocity \cite{Li21APL_NonreciprocalSSH}, interface interaction \cite{Wang20NanoLett_interaction}(Fig.\ref{figs:fig2}g), disorder \cite{Lin20PRR_disorder}, nonlinear harmonic generation \cite{Kruk19NatNanotech_SSH,Yuan22LPR_SSH}, and polarization control \cite{Slobozhanyuk16LPR,Tripathi21NanoPhot}. Specifically, in \cite{Jean17NatPho_lasing}, the authors experimentally observed lasing in the topological edge states of a 1D lattice of coupled polariton micropillars that implements an orbital version of the SSH model. Topological single-mode lasing in 1D SSH model has also been demonstrated in an array of coupled microring resonators \cite{Zhao18NC_lasing,Parto18PRL_lasing}. Furthermore, motivated by the high quality factors and small mode volumes of nanocavity, topological lasing has also been demonstrated in 1D SSH model based on nanocavity array \cite{Ota18ComPhy_lasing,Han19Light_lasing} (Fig.\ref{figs:fig2}b). The topological midgap states of the 1D SSH model under gain and loss were studied in \cite{Schomerus13OL_nonH,Poli15NC_nonH,Zeuner15PRL_nonH,Song19PRL_recover}(Fig.\ref{figs:fig2}c), where the authors in \cite{Zeuner15PRL_nonH} managed to monitor the topological transition by employing bulk dynamics only. Non-Hermitian skin effect, where all eigenstates of the system are localized at the system boundary, was studied in 1D SSH array of coupled resonant optical waveguides \cite{Zhu20PRR_skin,Lin21OE_squareroot}. Moreover, periodically bended waveguides could be exploited to mimic time-dependent driving, which allows the study of Floquet SSH models \cite{Cheng19PRL_drivenSSH,Petracek20PRA_drivenSSH}(Fig.\ref{figs:fig2}d). The SSH edge modes could also be used for quantum technological applications, such as, protection of biphoton states \cite{Redondo18Science_biphoton,Klauck21PhoRes_biphoton}(Fig.\ref{figs:fig2}e) or as quantum photonic baths for quantum simulations \cite{Saxena22ACSpho_bath}(Fig.\ref{figs:fig2}f). Strongly enhanced third-harmonic generation was recently observed in experiments using zigzag array of dielectric nanoparticles \cite{Kruk19NatNanotech_SSH} and silicon photonic crystal nanocavities \cite{Yuan22LPR_SSH}. Finally, due to the polarization-dependent edge states in zigzag SSH array, selective excitation of the topological edge states could be controlled by the handedness of the incident light as demonstrated in \cite{Slobozhanyuk16LPR}(Fig.\ref{figs:fig2}h) or for polarization control of photoluminescence of nanocrystals \cite{Tripathi21NanoPhot}.

%%%%%%%%%%%%%%%%%%%%%%%%%%%%
\section{Topological photonics in 2D}\label{sec4}
%%%%%%%%%%%%%%%%%%%%%%%%%%%

Photonic topological states in 2D in general could be classified as time-reversal symmetry breaking or preserving. For the photonic quantum Hall states with time-reversal symmetry broken by external magnetic fields, a topologically nontrivial bulk bandgap is characterized by the Chern number and the corresponding edge states within the bandgap exhibit chiral behavior, i.e., they can only propagate in one direction as the backpropagating states are completely removed by the magnetic fields. For this reason, the photonic quantum Hall states are absolutely robust against defects and disorders as long as the bandgap protecting these states persists. In the case of time-reversal symmetry preserving, several categories of photonic topological states with both the conventional bulk-edge correspondence, such as, quantum spin Hall states, quantum valley Hall states and the anomalous bulk-corner correspondence, such as, second-order topological corner states, have been explored in the literature. For the photonic quantum spin or valley Hall states, as their time-reversal partners alway exist, these states are only robust against certain types of defects and disorders, i.e., their robustness is weaker than that of the quantum Hall states. In the following, we will mainly discuss these four categories of photonic topological states in 2D.

\subsection{Photonic quantum Hall states}

The quantum Hall effect could be characterized by the Chern number defined by \cite{Lu14NatPho_review}
\begin{gather}
C_n=\frac{1}{2\pi}\iint_{\textrm{FBZ}} \mathbf{F}_n(\mathbf{k}) d^2\mathbf{k}
\end{gather}
where $\mathbf{F}_n(\mathbf{k})=\nabla_{\mathbf{k}} \times \mathbf{A}_n(\mathbf{k}) $ and $\mathbf{A}_n(\mathbf{k})=\langle \mathbf{u}_n(\mathbf{k}) | i \nabla_{\mathbf{k}} | \mathbf{u}_n(\mathbf{k}) \rangle$ are the Berry curvature and Berry connection respectively, and $\mathbf{u}_n(\mathbf{k})$ is the spatially periodic part of the Bloch function for the $n$-th band. Here, the integral in momentum space is performed over the first Brillouin zone (FBZ) and the inner product of the Berry connection is defined as $\langle \mathbf{u}_n | \mathbf{u}_m \rangle = \iint_{\textrm{UC}} \epsilon(\mathbf{r}) \mathbf{u}_n^*(\mathbf{r}) \mathbf{u}_m(\mathbf{r}) d^2\mathbf{r}$,
where  the integral in real space is performed over the unit cell (UC).  As the Berry curvature is odd under time-reversal operation, i.e., $\mathcal{T}: \mathbf{F}_n(-\mathbf{k}) \rightarrow  -\mathbf{F}_n(\mathbf{k})$, for a photonic system with time-reversal symmetry, the integral of the Berry curvature over the first Brillouin zone is alway zero (i.e., $C_n=0$). Different approaches to calculate the Chern number numerically for both Hermitian and non-Hermitian photonic crystals, e.g., based on Wilson-loop or Green's function, have been proposed recently \cite{Paz19AQT_tutorial,Wang19NJP_wilson,Wang20FOpto_comsol,Zhao20OE_chern,Prudencio20CP_chern_nonH,Chen21PRA_wilson_nonH}.

\begin{figure*}
\includegraphics[width=0.9\textwidth]{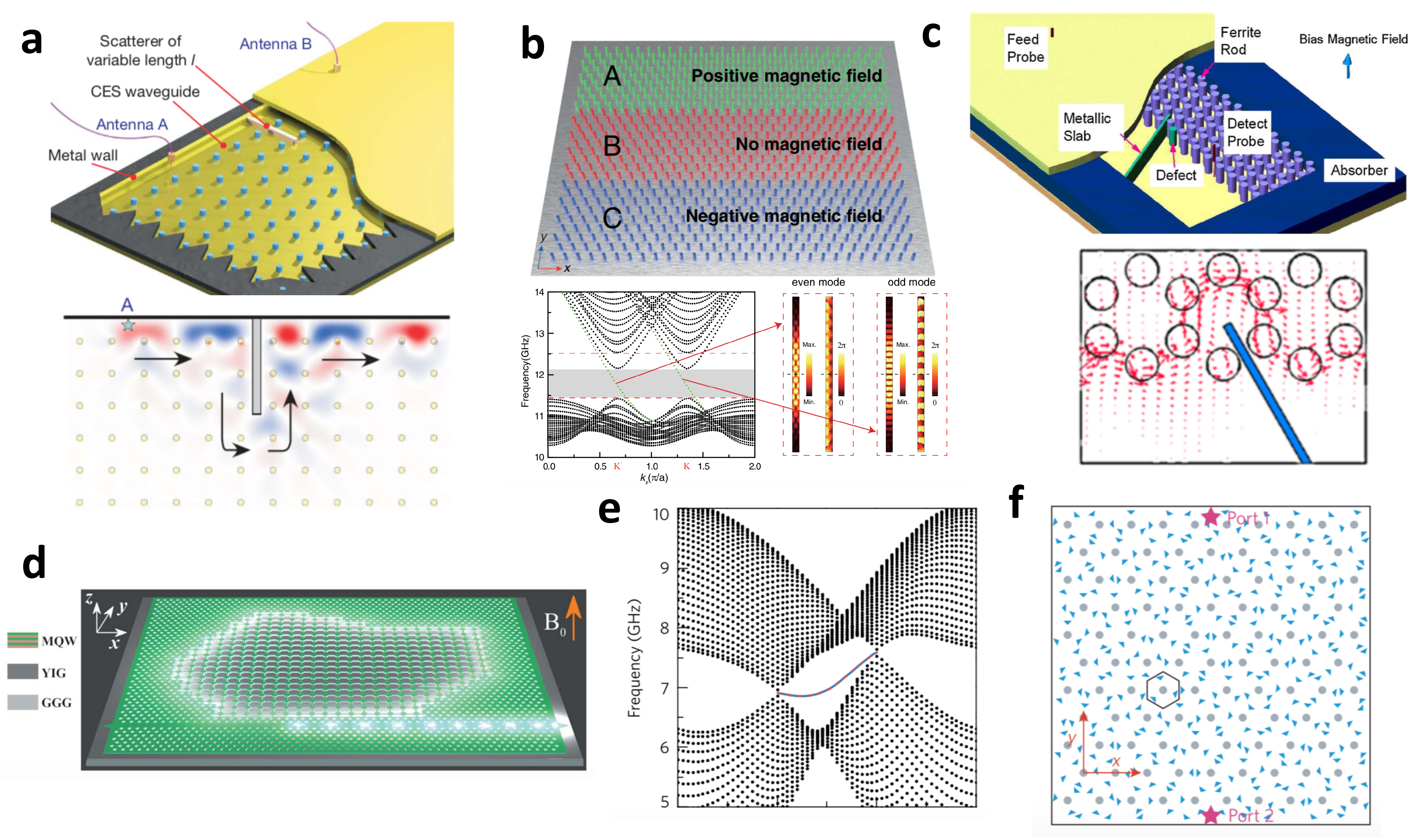}
\caption{Photonic quantum Hall states and their applications. (a) Experimental observation of one-way topological electromagnetic edge states. Reproduced with permission from~\cite{Wang09nature}, Copyright (2009) by Springer Nature. (b) One-way large-area topological waveguide states in magnetic photonic crystals. Reproduced with permission from~\cite{Wang21PRL_LargeArea}, Copyright (2021) by the American Physical Society. (c) Self-guiding electromagnetic edge states along an edge bounded by air. Reproduced with permission from~\cite{Poo11PRL}, Copyright (2011) by the American Physical Society. (d) Nonreciprocal lasing in topological cavities of arbitrary geometries. Reproduced with permission from~\cite{Bahari17Science}, Copyright (2017) by the American Association for the Advancement of Science. (e) Experimental observation of photonic antichiral edge states. Reproduced with permission from~\cite{Zhou20PRL_antichiral}, Copyright (2020) by the American Physical Society. (f) Experimental observation of topological Anderson insulator in disordered gyromagnetic photonic crystals. Reproduced with permission from~\cite{Liu20PRL_disorder}, Copyright (2020) by the American Physical Society. }
\label{figs:fig3}
\end{figure*}

The first photonic Chern insulator in photonic crystals with broken time-reversal symmetry was proposed by Haldane and Raghu \cite{Haldane08PRL} in 2008 in an effort to construct analogs of quantum Hall edge states for electromagnetic waves. They demonstrated that the Dirac points in the transverse electric (TE) photon bands of a hexagonal 2D array of cylindrical dielectric rods could be gapped out by adding an imaginary off-diagonal component to the permittivity tensor.  As a result, the split bands acquire non-zero Chern numbers of $\pm1$ and unidirectional photonic modes with no possibility of being backscattered at bends or imperfections emerge along an interface between two magneto-optic photonic crystals with different topological properties. Later, Wang et al, \cite{Wang08PRL_QH} pointed out that Dirac points in the band structure are not strictly necessary for observing these reflection-free one-way edge modes and they demonstrated the existence of transverse magnetic (TM) photonic bandgaps with nonzero Chern number in a square-lattice gyromagnetic photonic crystal operating at microwave frequencies. This theoretical prediction was soon experimentally verified by the same authors in \cite{Wang09nature} using an interface between a gyromagnetic photonic-crystal slab and a metal wall as shown in Fig.\ref{figs:fig3}a, where even large metallic scatterers placed in the path of the propagating edge modes do not induce reflections.

While the confining edge in \cite{Wang09nature} was constructed by cladding the photonic crystal with a metal wall, different methods to construct a confining edge for chiral edge states were studied. In \cite{Fu10APL_QH}, a one-way waveguide formed between a gyromagnetic photonic crystal and a normal dielectric photonic crystal was experimentally demonstrated and the authors found that when the waveguide width changes, while the  forward modes are very robust against intrusion of a metal plate,  the backward-propagating waves are very sensitive to the width of the waveguide. The waveguide at the boundary of two adjacent magneto-optical photonic crystals with opposite magnetic biases was studied in \cite{Lai20AIP} and the authors found that this waveguide can simultaneously support symmetrical and anti-symmetrical topological edge states, which possess the same direction of energy propagation, however, their directions of phase propagation are opposite. Furthermore, by inserting a domain of an ordinary photonic crystal sandwiched between two domains of magnetic photonic crystals, the authors in \cite{Wang21PRL_LargeArea} experimentally demonstrated the existence of large-area one-way waveguide states with amplitude uniformly distributed over the nonmagnetized domain (see Fig.\ref{figs:fig3}b). Interestingly, the possibility to have one-way edge states at the edge of a single gyrotropic photonic crystal bounded by air due to the light-cone effect was also studied in \cite{Ochiai09PRB,Lin09PRB,Poo11PRL,Tasolamprou21PRapp,Liu12OL_slab}. For example, in \cite{Poo11PRL} (see Fig.\ref{figs:fig3}c), self-guiding electromagnetic edge states outside the light cone along the zigzag edge of a honeycomb magnetic photonic crystal without requiring an ancillary cladding layer was experimentally demonstrated. Recently, by bringing the bandgap that supports one-way edge states below the light-cone, self-guiding electromagnetic edge states were also shown to exist in a square photonic crystal bounded by air \cite{Tasolamprou21PRapp}. One-way electromagnetic modes could further be sustained by the edge of a gyromagnetic photonic crystal slab when the photonic bandgap supporting the chiral modes is below the light-cone \cite{Liu12OL_slab}. At infrared and terahertz frequencies, graphene placed in a static magnetic field can be characterized as an electrically gyrotropic material. As such, by periodically patterning monolayer graphene with nanoholes, topological one-way plasmonic edge states at deep-subwavelength scale operable up to infrared frequencies have been demonstrated \cite{Jin17PRL_graphene,Pan17NC_graphene}.

The one-way electromagnetic edge states have found a wide range of applications \cite{Wang22FronMater}, such as, one-way cross-waveguide splitter \cite{He10APL}, unidirectional channel-drop filter \cite{Fu11APL_filter}, one-way waveguide with large Chern numbers \cite{Skirlo14PRL,Skirlo15PRL}, dual-topology induced light-trapping in lower dimensions \cite{Li18NC_dislocation}, observation of unpaired photonic Dirac point \cite{Liu20NC_unpaired}, nonreciprocal lasing \cite{Bahari17Science} (Fig.\ref{figs:fig3}d), antichiral edge states \cite{Chen20PRB_Antichiral, Zhou20PRL_antichiral} (Fig.\ref{figs:fig3}e), nonreciprocal Goos-H\"anchen shift \cite{Ma20OE_shift}, and nonlinear frequency mixing \cite{You20SciAdv,Lan20PRB_nonlinear}. Because backscattering in conventional optical devices or systems dominates over all other loss mechanisms as the group velocity of electromagnetic waves becomes small, the one-way electromagnetic edge states without backscattering provide interesting opportunities for slow-light related applications \cite{Yang13APL_slow,Chen19PRB_slow,Chen19PhoRes_slow,Chen20OL_slow,Zhuang21OE_slow}. For example, in \cite{Yang13APL_slow}, group velocity more than one order of magnitude less than the speed of light was experimentally demonstrated in a waveguide sandwiched between a gyromagnetic photonic crystal and a metal cladding. In \cite{Chen19PRB_slow}, a waveguide composed of two magneto-optical photonic crystals with the same Chern number was shown to provide a mechanism to create a unique group-dispersionless slow-light state due to the strong interaction between the two counterpropagating one-way edge states in the two composite semi-infinite magneto-optical photonic crystals. This slow-light state could be used to realize a switchable slow light rainbow trapping \cite{Chen19PhoRes_slow}, where different frequency components of a wave packet are separated and stored at different positions. Apart from the strong interaction between the two counterpropagating one-way edge states studied in \cite{Chen19PRB_slow}, topological slow-light state could also be achieved by strong interactions between two regular co-propagating one-way edge states in a single-channel \cite{Chen20OL_slow} or double-channel waveguide \cite{Zhuang21OE_slow}.

While the one-way electromagnetic edge states are robust against small disorder, the fate of these states under strong disorder has also been studied \cite{Mansha17PRB_disorder,Yang19PRB_disorder,Yang20OE_disorder,Liu20PRL_disorder,Zhou20Light_disorder}. Amorphous analogs of 2D photonic Chern insulators consisting of gyromagnetic rods with only short-range order but no long-range order were studied in \cite{Mansha17PRB_disorder}, where the authors found that nonreciprocal transmission exists even at very low levels of short-range order with no discernible spectral gaps. Amorphous magnetic photonic lattices with only short-range order were also studied in \cite{Yang19PRB_disorder}, in which single-mode and multimode topological edge states can exist despite the amorphous nature of the lattices. Disorder can induce a topological transition from a trivial insulator to a nontrivial insulator, i.e., a topological Anderson insulator. Recently, the authors in \cite{Liu20PRL_disorder} experimentally demonstrated a photonic topological Anderson insulator in a 2D disordered gyromagnetic photonic crystal as shown in Fig.\ref{figs:fig3}f, where they directly observed the disorder-induced topological phase transition from a trivial insulator to a topological Anderson insulator with robust chiral edge states. By gradually deforming the amorphous lattice of a photonic Chern insulator into a liquid-like lattice through the glass transition, the authors in \cite{Zhou20Light_disorder} experimentally observed the closing of the mobility gap and the disappearance of the topological edge states.

%%%%

\subsection{Photonic quantum spin Hall states}

In the quantum spin Hall effect of electronic materials that preserve the time-reversal symmetry, electrons of spin-up and spin-down experience the opposite magnetic fields, and as a result, the spin-up and spin-down electrons move along the opposite directions of a system edge, forming the so-called spin-momentum locked helical transportation. This effect could be characterized by a topological invariant, called the spin Chern number,
\begin{gather}
C_{\text{spin}}=\frac{1}{2}(C_{\uparrow}-C_{\downarrow})
\end{gather}
where $C_{\uparrow}$ and $C_{\downarrow}$ are the Chern numbers of the spin-up and spin-down electrons, respectively.

\begin{figure*}
\includegraphics[width=0.9\textwidth]{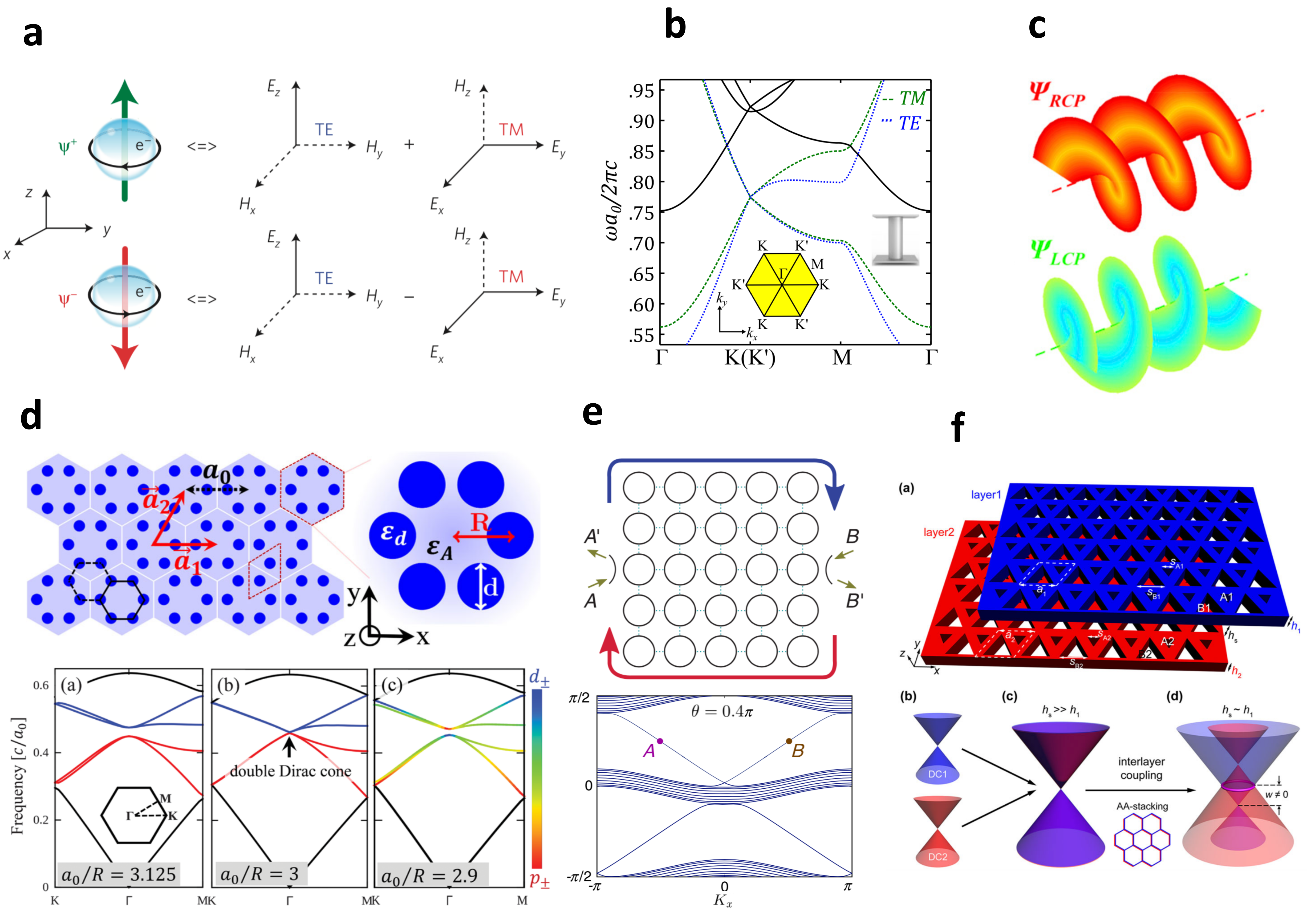}
\caption{Different mechanisms to emulate the two spin states of electrons using electromagnetic waves for the realization of photonic quantum spin Hall insulators. (a) Based on TE+TM/TE-TM. Reproduced with permission from~\cite{Khanikaev13NatMat}, Copyright (2013) by Springer Nature. (b) Based on TE/TM. Reproduced with permission from~\cite{Ma15PRL}, Copyright (2015) by the American Physical Society. (c) Based on TE+$i$TM/TE-$i$TM. Reproduced with permission from~\cite{He16PNAS_spin}, Copyright (2016), under CC BY-NC-ND or CC BY. (d) Based on double TM Dirac cones. Reproduced with permission from~\cite{WuHu15PRL}, Copyright (2015) by the American Physical Society. (e) Based on the clockwise/anti-clockwise whispering gallery modes of light in a ring resonator. Reproduced with permission from~\cite{LiangChong13PRL}, Copyright (2013) by the American Physical Society. (f) Based on the layer pseudospin in a two-layer photonic slab system. Reproduced with permission from~\cite{Chen19LPR_layer}, Copyright (2019) by John Wiley and Sons. }
\label{figs:fig4}
\end{figure*}

\begin{figure*}
\includegraphics[width=0.9\textwidth]{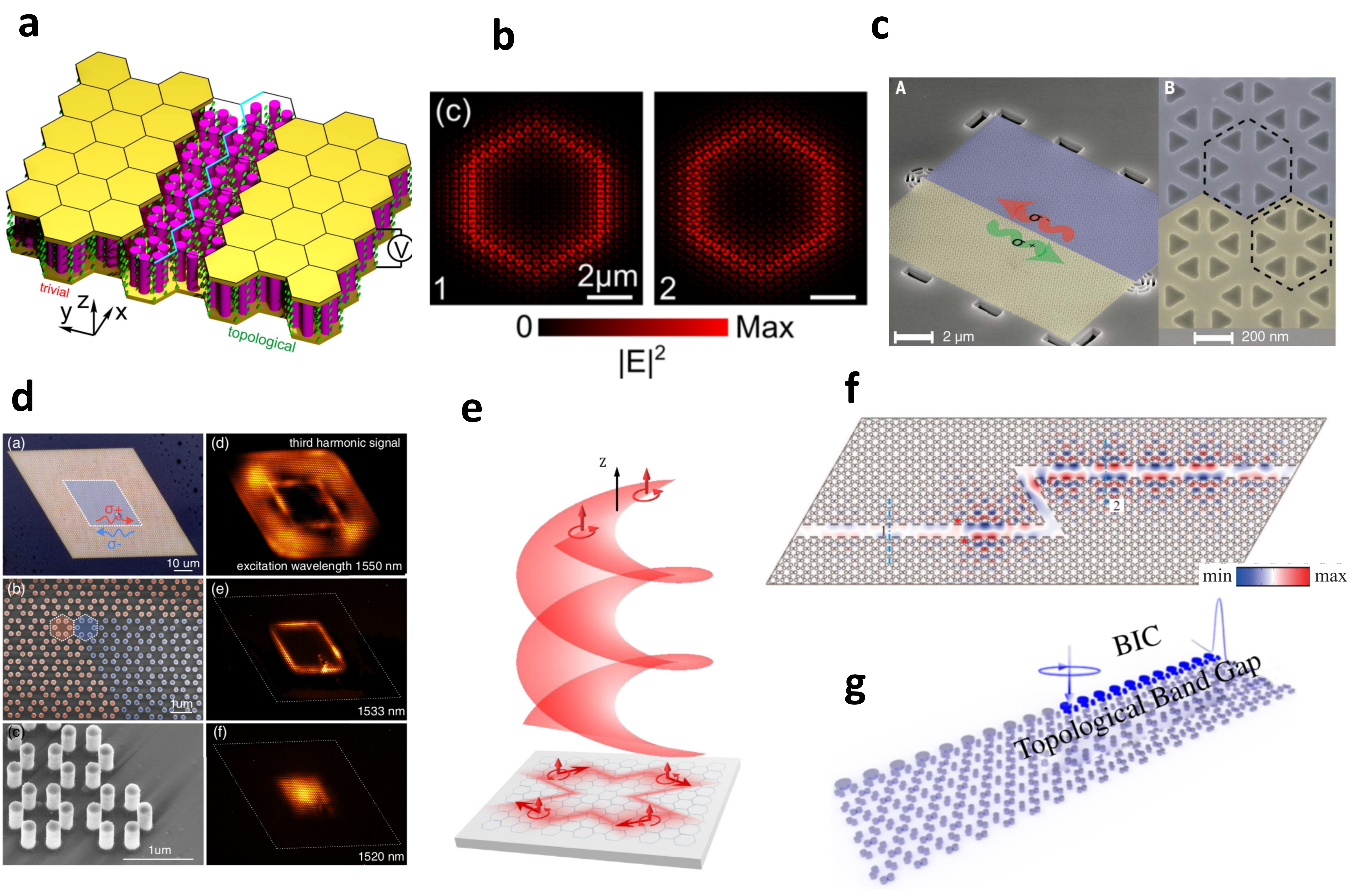}
\caption{Different applications of helical edge states based on the proposal in \cite{WuHu15PRL}. (a) Reconfigurable topological photonic crystal. Reproduced with permission from~\cite{Shalaev18NJP_reconfig}, Copyright (2018), under CC BY 3.0. (b) Topological photonic ring resonator. Reproduced with permission from~\cite{Mehrabad20APL_cavity}, Copyright (2020) by AIP Publishing. (c) Chiral coupling between the helical topological edge modes and a quantum emitter. Reproduced with permission from~\cite{Barik18Science}, Copyright (2018) by the American Association for the Advancement of Science. (d) Topological imaging of bulk and edge states via nonlinear third harmonic generation. Reproduced with permission from~\cite{Smirnova19PRL_THG}, Copyright (2019) by the American Physical Society. (e) Topological vortex laser based on the out-of-plane radiation feature of spin-momentum locking. Reproduced with permission from~\cite{Yang20PRL_laser}, Copyright (2020) by the American Physical Society. (f) Pseudospin-polarized topological line defect robust against a bending gap. Reproduced with permission from~\cite{Chen20IEEE_line}, Copyright (2020) by IEEE. (g) Bound topological edge states in the continuum exhibiting pseudospin-momentum locking unidirectional propagation at the air edge of a nontrivial expanded domain. Reproduced with permission from~\cite{Zhang21PRapp_BIC}, Copyright (2021) by the American Physical Society. }
\label{figs:fig5}
\end{figure*}

As electromagnetism has two components of electricity and magnetism, this provides many different ways to mimic the two spins of electrons. In \cite{Khanikaev13NatMat}, the linear combinations of TE and TM modes in 2D, i.e., TE+TM and TE-TM (see Fig.\ref{figs:fig4}a) are used to emulate the two spin states of electrons when the permittivity and permeability are tuned to be equal. Furthermore, the magneto-electro coupling could be introduced through the bi-anisotropic response and as such, the system proposed provides an exact emulation of the Kane-Mele Hamiltonian of electronic topological insulators, where one-way spin-polarized transport of photonic edge states robust against different types of defects was theoretically demonstrated. Since bi-anisotropic coupling in metamaterials usually is weak and highly dispersive, the permittivity and permeability matching condition can only be satisfied in a narrow frequency range. In \cite{Chen14NC_uniaxial}, the authors proposed a different approach to realize the coupling of TE and TM modes by confining the modes to a waveguide via symmetry reduction. Moreover, by measuring the magnitude and phase of the fields, gapless spin-filtered edge states were successfully observed in experiments. Later, degenerate TE and TM Dirac cones overlapping at the K/K' point of the Brillouin zone (see Fig.\ref{figs:fig4}b) were proposed to emulate the two spin states of electrons in a parallel-plate metal waveguide \cite{Ma15PRL} filled with a periodically arranged hexagonal array of metallic cylinders connected to the top and/or bottom metal plates. In this setup, a finite bianisotropy could be generated by a finite vacuum gap between the rods and one of the metal plates to effectively emulate spin-orbit interaction. Through first-principles electromagnetic simulations, the authors demonstrated that topologically protected surface waves could be guided without reflections along sharp bends. This idea was experimentally studied in \cite{Lai16SciRep_delayline} for the realization of a reflections-free compact delay line and in \cite{Cheng16NatMat} for the realization of a reconfigurable photonic topological insulator where the electromagnetic propagation pathways could be controlled along any desired path.  Extensions of the idea based on degenerate TE and TM double Dirac cones to a metasurface made of dielectric disks \cite{Slobozhanyuk19APL} and a metallic dual-metasurface \cite{Bisharat19LPR} were experimentally demonstrated with the corresponding spin-momentum locked edge states observed by the method of near-field imaging.

Double Dirac cones due to accidental degeneracy between TE and TM modes could also be achieved in photonic crystals with anisotropic permittivity \cite{Chen18LPR_Accidental} and with nonzero bianisotropy, topological bandgap supporting robust transport of gapless edge states was theoretically demonstrated. Additionally, a pair of double-degenerate TE and TM modes in a gyrotropic photonic crystal with both gyroelectric and gyromagnetic responses, where off-diagonal terms in permittivity and permeability tensors coexist, was employed to realize a photonic topological insulator in \cite{Sun19Crystals}. Apart from the combinations of TE and TM modes \cite{Ma15PRL,Chen18LPR_Accidental,Sun19Crystals}, TE+TM/TE-TM \cite{Khanikaev13NatMat}, in \cite{He16PNAS_spin}, the left circular polarization and right circular polarization (see Fig.\ref{figs:fig4}c), i.e., the combinations of TE+$i$TM/TE-$i$TM were proposed to mimic the electronic spin states and thus to realize photonic topological insulators. In 2015, Wu and Hu \cite{WuHu15PRL} proposed a photonic topological insulator based on purely the TM modes in an all-dielectric photonic crystal by deforming a honeycomb lattice of cylinders into a triangular lattice of cylinder hexagons (see Fig.\ref{figs:fig4}d). This achievement is based on the fact that there are two 2D irreducible representations of the $C_6$ symmetry group and thus a double-degenerate Dirac cone could be realized by the band folding mechanism in the Brillouin zone. Furthermore, a band inversion could be simply achieved by expanding or shrinking the six cylinders within the hexagon and helical edge states with the pseudospin mimicked by the angular momentum of the TM modes were theoretically demonstrated. Later, an experiment \cite{Yang18PRL_spinExp} based on a photonic crystal made of Al$_2$O$_3$ cylinders confirmed this proposal in the microwave regime. Moreover, pseudospin-up and pseudospin-down could also be emulated by the two directions of propagation of light in a  ring resonator \cite{LiangChong13PRL} (see Fig.\ref{figs:fig4}e) or the layer degree of freedom in two-layer photonic systems \cite{Chen19LPR_layer,Wu19AOM_layer} (see Fig.\ref{figs:fig4}f).

Due to the simple design strategy based on symmetry consideration, the proposal by Wu and Hu \cite{WuHu15PRL} has attracted a great interest in the community, e.g., it was demonstrated that the idea could also be applied to dielectric slab with holes \cite{Barik16NJP_hole,Anderson17OE_hole} rather than the original cylinders-in-air configuration as well as applied to exciton-polaritons \cite{Liu20Science_polarition,Li21NC_polariton} and the topological helical edge states have been further observed at both telecom \cite{Gorlach18NC_far,Parappurath20SciAdv} and visible \cite{Peng19PRL_visible,Liu20NanoLett_Z2} wavelengths. The idea has also found a wide range of applications, e.g., in realizing reconfigurable topological photonic crystals with tunable edge states \cite{Shalaev18NJP_reconfig,Cao19SciBlu,Wang19JAP_reconfig} (Fig.\ref{figs:fig5}a), topological ring-cavity and whispering gallery modes \cite{Yang18OE_cavity,Mehrabad20APL_cavity,Sun21PRB_cavity} (Fig.\ref{figs:fig5}b), chiral coupling between the helical topological edge modes and quantum emitters \cite{Barik18Science}, topological all-optical logic gates \cite{He19OE_logic}, topological nonlinear third-harmonic generation \cite{Smirnova19PRL_THG} and novel topological lasing behaviors \cite{Shao20NatNanoTech_laser,Yang20PRL_laser}. For examples, in \cite{Barik18Science}, the chiral emission of a quantum emitter into the counterpropagating helical edge modes was observed experimentally (see Fig.\ref{figs:fig5}c) and in \cite{He19OE_logic}, topological filter and all-optical logic gates were designed based on the robust transport of helical edge states. Imaging topological edge states via nonlinear optical process may offer superior contrast, sensitivity, and large imaging area, which was demonstrated in \cite{Smirnova19PRL_THG}, where the authors experimentally observed strong third-harmonic generation and demonstrated that variation of the pump-beam wavelength enables independent high-contrast imaging of either bulk modes or spin-momentum-locked edge states (see Fig.\ref{figs:fig5}d). Moreover, the band inversion between the shrunken and expanded structures provides a novel mode confinement mechanism due to the opposite parities of the bulk wavefunctions in trivial and nontrivial photonic crystals. Based on this principle, a topological bulk laser exhibiting single-mode lasing with vertical emission directionality was experimentally realized in \cite{Shao20NatNanoTech_laser}. Later on, the authors \cite{Yang20PRL_laser}, exploiting the out-of-plane radiation feature of spin-momentum locking (see Fig.\ref{figs:fig5}e), demonstrated a high performance topological vortex laser and found that the near-field spin and orbital angular momentum of the topological edge mode lasing has a one-to-one far-field radiation correspondence.

While the typical boundary for topological edge states based on the proposal in \cite{WuHu15PRL} is constructed between shrunken and expanded domains, other methods to construct the boundary for observing topological edge states were also proposed and studied in the literature \cite{Gao19AO_defect,Chen20IEEE_line,Yang22OE_finiteW,Zhang21PRapp_BIC}, e.g., in \cite{Chen20IEEE_line}, a line defect was introduced directly into a single nontrivial expanded domain and pseudospin-polarized transport of electromagnetic waves that can bypass a bending gap was demonstrated (see Fig.\ref{figs:fig5}f). The physics of this unusual phenomenon is due to the strong coupling of the topological modes located at the up and down boundaries of the line defect. One can further replace the air line defect by a trivial shrunken domain with finite width and study the coupling of the topological edge modes when changing the width of the trivial domain as shown in \cite{Yang22OE_finiteW}, where pseudospin-preserving and pseudospin-flipping processes were observed in microwave experiments. One can even consider the air edge of a nontrivial domain \cite{Zhang21PRapp_BIC}, where bound topological edge states in the continuum exhibiting topological features inherited from the nontrivial topology of bulk photonic bands, such as spin-momentum locking unidirectional propagation (see Fig.\ref{figs:fig5}g), were demonstrated in microwave experiments.

%%%%%

\subsection{Photonic quantum valley Hall states}

\begin{figure}
\includegraphics[width=0.5\textwidth]{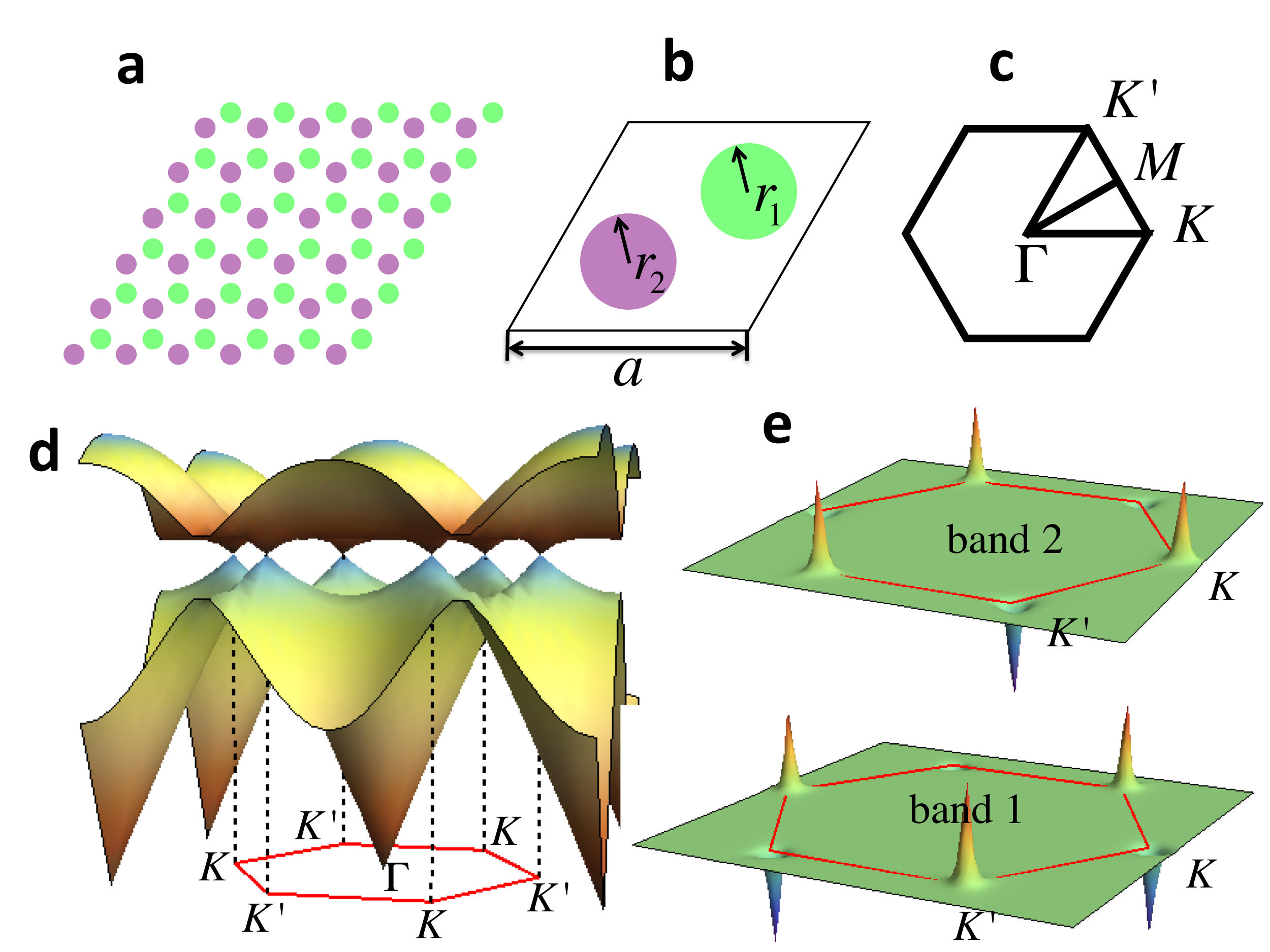}
\caption{Schematics to illustrate the idea of valley Chern number. (a) A honeycomb photonic crystal with two different kinds of dielectric cylinders. (b) The unit cell and (c) the first Brillouin zone of the photonic crystal. (d) When $r_1=r_2$, the first two bands of the TM modes form Dirac cones around $K/K'$. (e) When $r_1\neq r_2$ with a small radius difference, the Dirac cones at $K/K'$ are gapped out and the resulting Berry curvature has sharp peaks around $K/K'$, whose integral around $K/K'$ gives the valley Chern number $C_{K/K'}=\pm 1/2$.  }
\label{figs:fig6}
\end{figure}

In photonic systems with time-reversal symmetry, the integral of the Berry curvature over the first Brillouin zone is zero as $\mathbf{F}_n(-\mathbf{k})=-\mathbf{F}_n(\mathbf{k})$. Nonetheless, for a photonic crystal with hexagonal symmetry, the Berry curvature could have nontrivial distributions around the $K/K'$ points (i.e., valleys) of the first Brillouin zone, which allows the definition of valley Chern number by integrating the Berry curvature around $K/K'$,

\begin{gather}
C_{K/K'}=\frac{1}{2\pi}\iint_{K/K'} \mathbf{F}_n(\mathbf{k}) d^2\mathbf{k}
\end{gather}

\begin{figure*}
\includegraphics[width=0.9\textwidth]{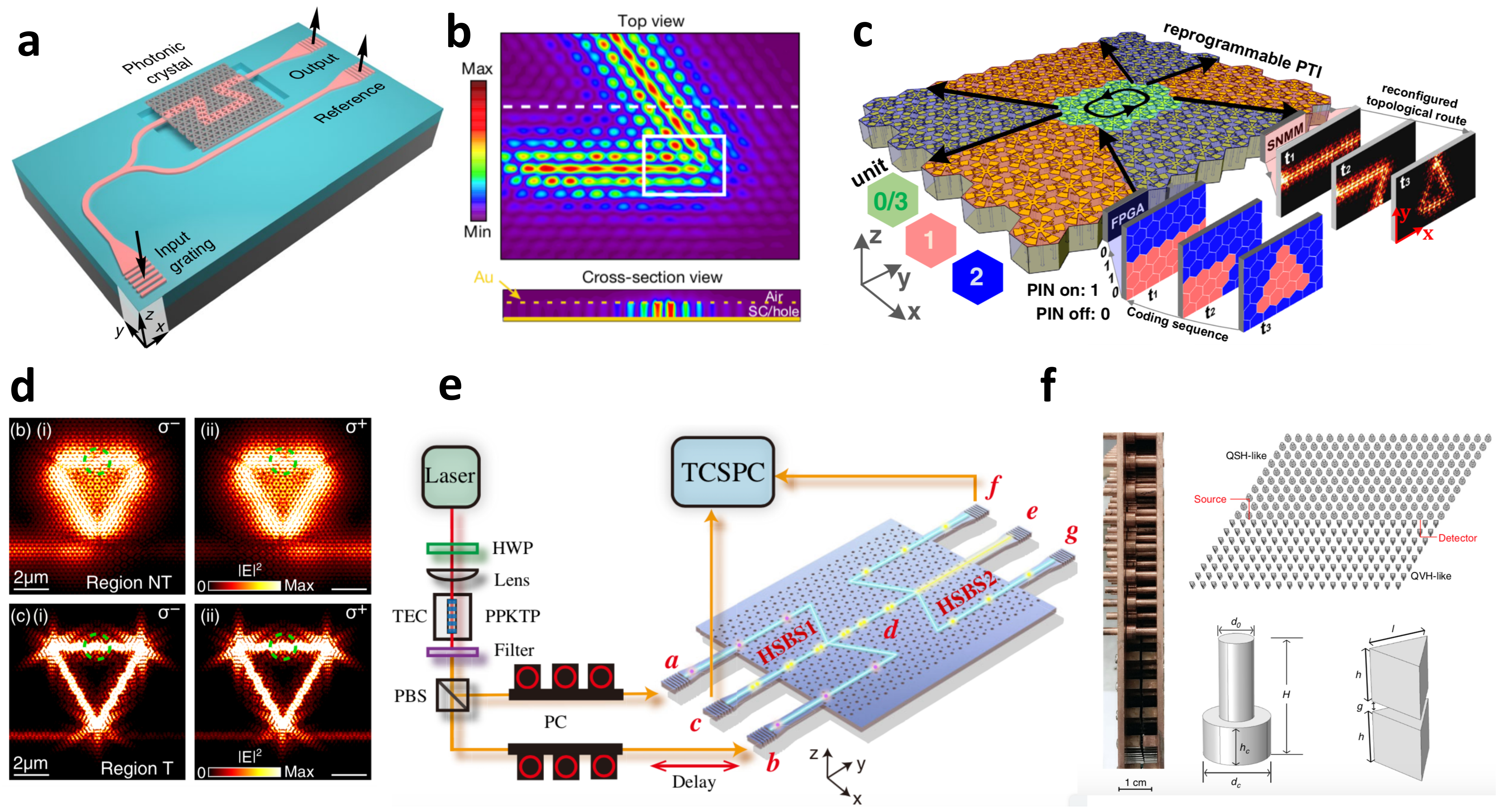}
\caption{Valley edge modes and their applications. (a) Schematic of a valley photonic crystal operating at telecommunication wavelengths consisting of a honeycomb lattice of two inverted equilateral triangular air holes per unit cell. Reproduced with permission from~\cite{Shalaev19NatNanotech}, Copyright (2019) by Springer Nature. (b) Electrically pumped topological laser with valley edge modes. Reproduced with permission from~\cite{Zeng20Nature_laser}, Copyright (2020) by Springer Nature. (c) A reprogrammable plasmonic topological insulator with nanosecond-level switching time. Reproduced with permission from~\cite{You21NC_reconfig}, Copyright (2021), under CC BY 4.0. (d) Chiral coupling between quantum emitters and valley edge modes. Reproduced with permission from~\cite{Mehrabad20Optica_dot}, Copyright (2020), under CC BY 4.0. (e) On-chip Hong-Ou-Mandel interference based on valley-dependent quantum circuits. Reproduced with permission from~\cite{Chen21PRL_quantum}, Copyright (2021) by the American Physical Society. (f) Spin-valley coupled edge states. Reproduced with permission from~\cite{Kang18NC_spinvalley}, Copyright (2018), under CC BY 4.0.}
\label{figs:fig7}
\end{figure*}

Note that in the literature, valley Chern number could also be referred to $C_v=C_K-C_K'$. To illustrate the idea, we consider a honeycomb photonic crystal with two cylinders in each unit cell \cite{Lan21PRA_SHG} (see Fig. ~\ref{figs:fig6}a-b). When the radii of the two cylinders in the unit cell are equal (assuming they have the same dielectric constant), the first two bands of the TM modes in the band structure form Dirac cones around $K/K'$ (see Fig. ~\ref{figs:fig6}c-d). One can break the inversion symmetry by considering the two cylinders with different radii and as such, the Dirac cones around $K/K'$ are gapped out, resulting in nontrivial Berry curvature distributions around  $K/K'$ (see Fig. ~\ref{figs:fig6}e). One can find  $C_{K/K'}=\pm 1/2$ for the two valleys $K/K'$ if the radius difference is small. To construct a domain wall between two valley photonic crystals with different topological properties, one can consider one photonic crystal (I) that is inversion symmetric to the other (II) and consequently, the two valleys will be transformed into each other in the two domains, i.e., $K_I/K'_{II}=K'_{II}/K_I$ and $C^I_{K/K'}=-C^{II}_{K/K'}=\pm 1/2$. So the difference of the valley Chern number across the domain wall interface is $C^I_{K/K'}-C^{II}_{K/K'}=\pm 1$, which means that at one valley there exists an interface mode with positive group velocity, whereas another one with negative group velocity exists at the other valley. This principle to generate nontrivial valley Chern number and the corresponding valley interface modes has been frequently used in the literature \cite{Baile21AdvPhoRes_review,Dong21AdvPhyX_review}. Apart from this common interface separating two valley photonic crystals, other possible ways to construct the interface, e.g., air edge \cite{Chen20OE_air,Feng22OL_valleyBIC}, or three-layer sandwich structures \cite{He20OE_3layer,Chen21ACSpho_3layer}, have also been studied. Recently, the concept of large valley Chern number has been proposed \cite{Xi20PhoRes_large,Yan21arxiv_large}, e.g., in \cite{Xi20PhoRes_large}, the authors considered a 2D photonic crystal made of hexamers of six dielectric rods in each unit cell and showed that by shrinking or expanding one set of rods in the hexamer,  a valley phase transition with emergent multiple edge states from the valley Chern number of $C_K=1/2$ to $C_K=3/2$ could be achieved. Furthermore, large valley Chern number has also been realized in two different frequency bands \cite{Yan21arxiv_large}. In the following, we briefly review different lattice structures, system platforms, and various applications of valley photonic crystals.

Valley photonic crystals have been studied in triangular lattices \cite{MaShvets16NJP,Ye17APL_microwave,Zhang19LPR_Circuitry,Dubrovkin20APL,Li20PRL_spinvalley}, honeycomb lattices \cite{Dong17NatMat,Chen17PRB_contrast,Yang18SciRep,Chen18PRapp_cylinder,Noh18PRL_waveguide,He19NC_slab,Shalaev19NatNanotech,Shalaev19Optica,Ma19LPR_kink,Yang20NatPho,Arora21Light}, and kagome lattices \cite{Deng19NanoPhot_kagome,Wong20PRRes_kagome}, where the unit cell contains one, two and three lattice sites, respectively. For the triangular lattice, Ma and Shvets \cite{MaShvets16NJP} theoretically proposed to use the TE modes of a triangular photonic crystal, where each unit cell contains a single triangular Si rod or its perturbed structure without inversion symmetry, to create the valley degrees of freedom. Following this proposal, different experimental implementations have been demonstrated. For examples, in \cite{Ye17APL_microwave}, a valley photonic crystal constructed by a triangular array of Y-shaped aluminum rods in the microwave regime was demonstrated by rotating the Y-shaped rods within the unit cells. In \cite{Zhang19LPR_Circuitry}, by rotating triangular scatterers within the unit cells, valley kink states at generic interfaces in subwavelength substrate-integrated photonic circuitry were experimentally demonstrated. Recently, valley edge modes at $\lambda=1.55\mu$m were experimentally observed in a photonic crystal using the TM modes based on a triangular air-hole design with broken inversion symmetry fabricated from a suspended slab geometry \cite{Dubrovkin20APL}. Honeycomb photonic crystals, where each unit cell contains two dielectric cylinders or air holes provide a convenient setup for inversion symmetry breaking by choosing the two cylinders or holes different. For example, valley photonic crystals based on dielectric cylinders in honeycomb lattices have been theoretically studied in \cite{Dong17NatMat,Chen17PRB_contrast,Yang18SciRep} and experimentally observed  at microwave \cite{Chen18PRapp_cylinder} and  at $\lambda=1450$nm based on array of evanescently coupled waveguides \cite{Noh18PRL_waveguide}. Motivated by on-chip integration, light transport based on valley edge modes in honeycomb lattices has also been experimentally demonstrated in the slab geometry with either circular holes \cite{He19NC_slab} or triangular holes \cite{Shalaev19NatNanotech,Shalaev19Optica,Ma19LPR_kink,Yang20NatPho,Arora21Light}(Fig.\ref{figs:fig7}a). Furthermore, valley edge modes could also be realized in kagome photonic crystals \cite{Deng19NanoPhot_kagome,Wong20PRRes_kagome}, where each unit cell contains three cylinders (or holes) and by expanding or shrinking the three cylinders within the unit cells,  nontrivial valley bandgap from gapped Dirac cones and corresponding valley edge modes could be created. Valley edge modes can also appear within different frequency bandgaps, achieving the so-called dual-band valley kink states \cite{Chen19AOM_dual,Tang20PRB_dual,Wei21NJP_dual}, which may find interesting applications, such as topologically protected second harmonic generation \cite{Lan21PRA_SHG}.

Valley edge modes have also been studied in plasmonic systems, e.g., in surface-wave photonic crystal on a single metal surface \cite{Gao17PRB_SW}, designer surface plasmon crystal comprising metallic patterns deposited on a dielectric substrate \cite{Wu17NC_designer}, graphene plasmonic crystals \cite{Qiu17OE_graphene,Jung18PRL_graphene,You20IEEE,Wang20OL_2layer}, metal nanoparticles \cite{Proctor20Nanophot_metal} and metal cylinders \cite{Saito21NanoLett_metal}. Valley edge modes have found a range of interesting applications, such as, lasing \cite{Zeng20Nature_laser,Noh20OL_laser,Gong20ACSPho_laser,Zhong20LPR_laser,Liu22OE_laser}, fiber \cite{Makwana20OE_fiber,Zhang21NanoPhot_fiber}, reconfigurable devices \cite{Wu18PRM_reconfig,Wang19NJP_reconfig,You21NC_reconfig}, slow light \cite{Yoshimi20OL_slow,Xie21PRApp_slow,Yoshimi21OE_slow,Arregui21PRL_slow}, chiral coupling to quantum emitters \cite{Yamaguchi19APE_dot,Barik20PRB_dot,Mehrabad20Optica_dot}, Mach-Zehnder interferometer \cite{Yang20JO_MZ}, on-chip quantum information processing \cite{Chen21PRL_quantum}, long-range deformations
 \cite{Xu20PRR_disorder}, and  logic gates \cite{Chao21JO_gate}. Especially, for topological lasing based on valley edge modes, recently, an electrically pumped terahertz quantum cascade laser was experimentally demonstrated in a triangular lattice of quasi-hexagonal holes drilled through the active medium of a terahertz quantum cascade laser wafer \cite{Zeng20Nature_laser} (see Fig.\ref{figs:fig7}b). Subsequently, single-mode lasing of valley-Hall ring cavities at telecommunication wavelength was experimentally realized in a structured and suspended membrane consisting of large and small holes \cite{Noh20OL_laser}. Reconfigurable topological states in valley photonic crystals were also proposed using BaTiO$_3$ \cite{Wu18PRM_reconfig} as well as liquid crystal \cite{Wang19NJP_reconfig} and recently, a reprogrammable plasmonic topological insulator, where the topological propagation route can be dynamically changed at nanosecond-level switching time, was demonstrated experimentally \cite{You21NC_reconfig} (see Fig.\ref{figs:fig7}c). Moreover, exploiting this ultrafast control feature, a topologically protected multi-channel optical analog-digital converter was also demonstrated. Finally, by combining the spin and valley degrees of freedom, spin-valley coupled edge states \cite{Gao18NatPhy_spinvalley,Kang18NC_spinvalley}(Fig.\ref{figs:fig7}f), coexistence of pseudospin- and valley-Hall-like edge states \cite{Chen20PRR_spinvalley}, as well as spin- and valley-polarized one-way Klein tunneling \cite{Ni18SciAdv_spinvalley}, have been proposed.

%%%%%%%%

\subsection{Second-order photonic topological corner states}

\begin{figure*}
\includegraphics[width=0.9\textwidth]{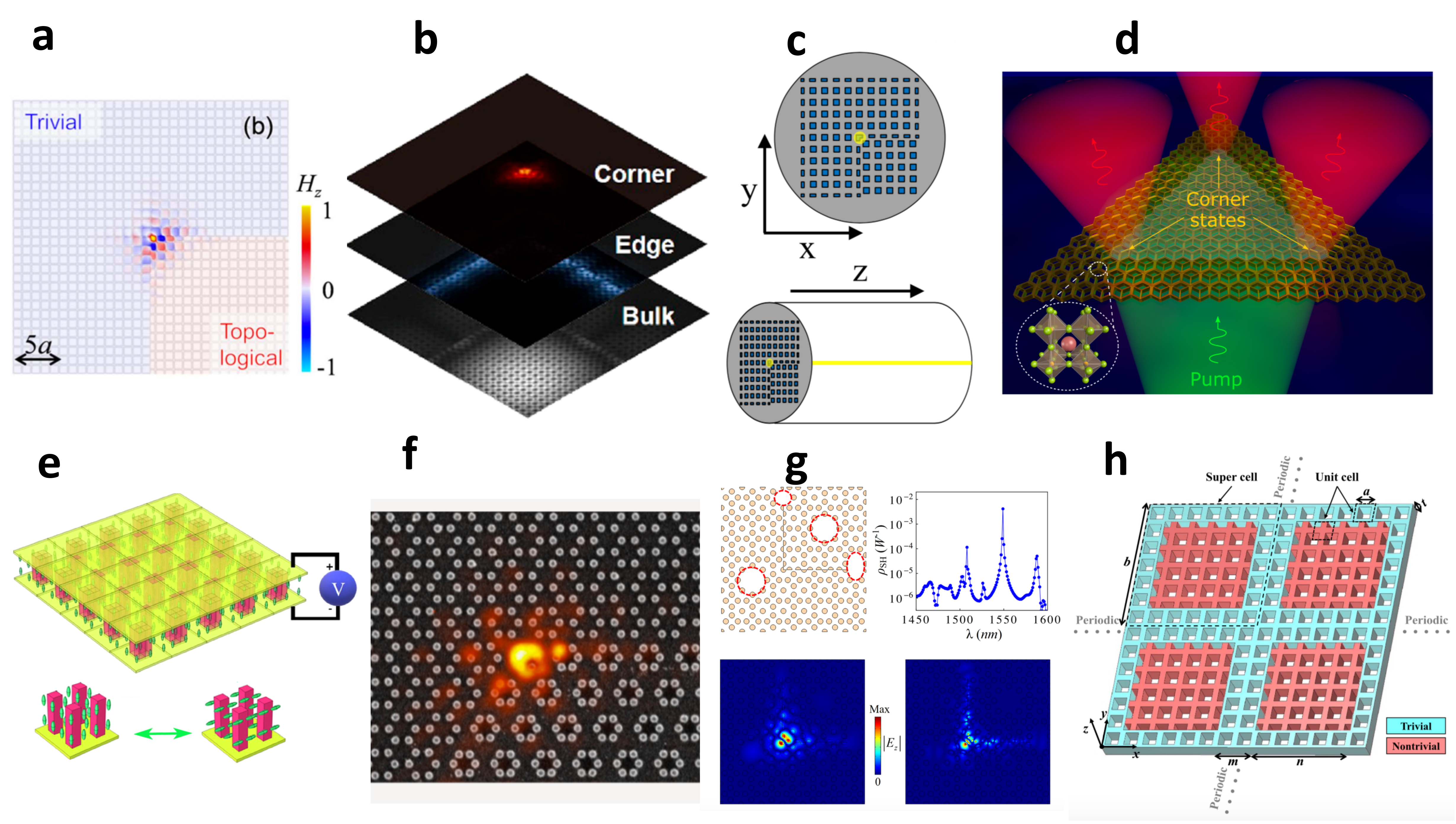}
\caption{Applications of topological corner states. (a) As nanocavity. Reproduced with permission from~\cite{Ota19Optica_cavity}, Copyright (2019), under CC BY 4.0. (b) Lasing of multidimensional topological states in a hierarchical scale via bulk, edge and corner states. Reproduced with permission from~\cite{Han20ACSpho_laser}, Copyright (2020) by American Chemical Society. (c) Topological photonic crystal fiber based on corner fiber mode. Reproduced with permission from~\cite{Gong21OL_fiber}, Copyright (2021) by Optica Publishing Group. (d) Enhanced photoluminescence of halide perovskite nanocrystals. Reproduced with permission from~\cite{Berestennikov21JPCC_PL}, Copyright (2021) by American Chemical Society. (e) Dynamically reconfigurable topological corner states. Reproduced with permission from~\cite{Hu21OL_reconfig}, Copyright (2021) by Optica Publishing Group. (f) Nonlinear imaging of nanoscale topological corner states. Reproduced with permission from~\cite{Kruk21NanoLett_image}, Copyright (2021) by American Chemical Society. (g) Topologically protected second harmonic generation via doubly resonant corner modes. Reproduced with permission from~\cite{Chen21PRB_SHG}, Copyright (2021) by the American Physical Society. (h) Coupled topological edge and corner modes in a photonic crystal slab consisting of a periodic array of SSH supercells. Reproduced with permission from~\cite{Zhang21JO_cornerarray}, Copyright (2021), under a CC BY license.}
\label{figs:fig8}
\end{figure*}

Topological corner states are higher-order topological phenomena where the topological boundary states appear in dimensions at least 2 orders lower than the bulk \cite{Kim20Light_review}. The existence of nontrivial bulk dipole or quadrupole moment could lead to the appearance of topological states localized at the corners of 2D photonic systems. Bulk dipole moment could be characterized by the 2D polarization $P=(P_x, P_y)$ given by \cite{Liu17PRL_zeroBerry},

\begin{gather}
P_i=\frac{1}{(2\pi)^2}\int_{\textrm{FBZ}} \textrm{Tr} [A_i,\mathbf{k}]d^2 \mathbf{k}
\end{gather}
with $[A_i,\mathbf{k}]^{mn}=i\langle u_{\mathbf{k}}^m|\partial_{k_i}|u_{\mathbf{k}}^n\rangle$ for $i=x,y$ and $m,n$ run over the occupied bands and $u^n_{\mathbf{k}}$ is the periodic Bloch function for the $n$th band. A direct integration of the Berry connection over the first Brillouin zone to get the polarization is numerically challenging. However, for photonic systems with inversion symmetry, the polarization could be obtained in a much simpler way by calculating the parities of the bands at high symmetry points of $\Gamma$ and X \cite{Liu17PRL_zeroBerry},

\begin{gather}
P_i=\frac{1}{2} \left( \sum_n q_i^n \hspace{0,2cm} \textrm{mod} \hspace{0,2cm} 2 \right), (-1)^{q_i^n}=\frac{\eta_n(X_i)}{\eta_n(\Gamma)}
\label{2Dpol}
\end{gather}
where $\eta_n$ represents the parity at the high symmetry points $\Gamma$ and $X$ of the first Brillouin zone for the $n$th band. Based on the above formula, a simple rule to judge whether a bandgap has nontrivial polarization could be deduced, i.e., an odd number of pairs of parities with opposite signs at $\Gamma$ and $X$ of the same band below the bandgap makes the bandgap topologically nontrivial whereas an even number implies the bandgap is trivial.

Most of the works on topological corner states from nontrivial 2D polarization in photonic systems are based on the 2D SSH model, i.e., each unit cell contains some cylinders and by expanding or shrinking the cylinders' positions away or towards the center of the unit cell, intra-cell and inter-cell hopping could be tuned, leading to a topologically trivial to nontrivial transition. For the square lattice 2D SSH model, the authors in \cite{Xie18PRB_corner} considered a 2D photonic crystal with four identical dielectric rods in each unit cell and by adjusting the distances between the nearby rods in the x and y directions, the emergence of edge and corner states can be controlled straightforwardly. The 2D square SSH model was experimentally realized in photonic crystal slabs with periodic dielectric rods on a perfect electric conductor \cite{Chen19PRL_obs} and in photonic crystals consisting of alumina cylinders sandwiched between two metallic plates \cite{Xie19PRL_vis}. The 2D square SSH model has also been studied with metallic nanoparticles \cite{Kim20NanoPhot_metal}, surface-wave photonic crystal consisting of metallic patterns on both sides of a dielectric substrate \cite{Zhang20AdvSci_surfacewave} and designer surface plasmon crystals composed of subwavelength metallic cylinders arranged in a 2D lattice structure on a metallic surface \cite{Wang21OE_THz}. Topologcial corner states have also been studied in honeycomb lattices with two cylinders in each unit cell \cite{Phan21OE_honey,KHO22pssb_honey}. In \cite{Phan21OE_honey}, the  authors found that topological corner states appear only for 60$^{\circ}$ corners, but absent for other corners, due to the sign flip of valley Chern number at the corner and in \cite{KHO22pssb_honey}, dual band topological corner states within both the first and third bandgaps were demonstrated. Kagome lattice whose unit cell contains three cylinders has also been studied for corner states. Especially, corner states in kagome lattices have been experimentally observed in waveguide arrays inscribed in glass samples using femtosecond laser technology \cite{Hassan19NatPho_kagome}, in array of dielectric cylinders arranged to form a kagome lattice between two parallel aluminium plates \cite{Li20NatPho_longrange}, and in metasurface fabricated on a silicon-on-insulator chip  consisting of  trimers of diamond-shaped holes \cite{Vakulenko21AdvMat_nearfield}.  Interestingly, in addition to the corner states due to nearest-neighbour interactions,  a new class of topological corner states induced by long-range interactions with a purely electromagnetic nature, which has no analogy in condensed-matter systems, exists in this lattice structure. Corner states have also been studied in a metal with air cavities forming kagome lattice \cite{Chen19OL_trunc}, in a plasmonic metasurface of metal nanoparticles arranged in a kagome lattice \cite{Proctor21APL_plasmon} and in an array of silicon rods forming a kagome lattice \cite{Shen21OE_kagome}.  By moving the dielectric rods continuously, the authors in \cite{Wang21PhoRes_C3} demonstrated that $C_3$ symmetric photonic crystals can switch between triangle and kagome lattice configurations, leading to rich higher-order topological phases and phase transitions. Finally, lattice structures where each unit cell contains six cylinders have also been explored \cite{Noh18NatPho_midgap,Proctor20PRR_robust,Wu21PhoRes_uncon,Gladstone22PRL_spincorner,KHO22OLT_6cylinder}, and some interesting features, such as, unconventional higher-order topology \cite{Wu21PhoRes_uncon} and spin-polarized fractional corner charges \cite{Gladstone22PRL_spincorner}, have been observed.

Corner states can serve as high-Q cavity modes, thus providing potential applications in enhancing light-matter interaction. For example, a corner state tightly localized in space with a high Q factor over 2000 was experimentally observed in \cite{Ota19Optica_cavity} (see Fig.\ref{figs:fig8}a), verifying its promise as a nanocavity. Further optimization of the Q factor of the corner state can bring it to 6000 \cite{Xie21OE_opt},  making the strong coupling to single quantum emitter possible. Indeed, the authors in \cite{Xie20LPR_QED} studied the coupling between single quantum dot and the corner state, where enhanced emission rate was observed when the quantum dot is on resonance with the corner state. The corner states could be pumped using in-plane excitation conditions as experimentally demonstrated in \cite{He21PhoRes_excitation} and have many interesting applications, such as, topological nanolasers \cite{Smirnova20Light_laser,Han20ACSpho_laser,Kim20NC_laser,Zhang20Light_laser} (Fig.\ref{figs:fig8}b), second-order topological photonic crystal fibers \cite{Gong21OL_fiber} (Fig.\ref{figs:fig8}c), high-quality optical hotspots \cite{Liu22ACSpho_hotspot}, enhancement of photoluminescence signal \cite{Berestennikov21JPCC_PL} (Fig.\ref{figs:fig8}d), rainbow trapping \cite{Liang22OL_rainbow}, dynamically reconfigurable topological corner states \cite{Hu21OL_reconfig} (Fig.\ref{figs:fig8}e), corner states in non-Hermitian photonic crystals \cite{Jiang22OL_nonH}. Corner states could also be exploited to enhance optical nonlinear effects, e.g., an interesting nonlinear imaging of nanoscale topological corner states through third harmonic generation was experimentally demonstrated in \cite{Kruk21NanoLett_image} (see Fig.\ref{figs:fig8}f) and enhanced second harmonic generation from a topological corner state and its directional out-of-plane emission were theoretically proposed in \cite{Guo21OE_SHG}. The corner states can exist in multiband gaps simultaneously \cite{Kim21AOM_multiband,Chen22PRApplied_multicorner}, providing the opportunity to realize highly efficient nonlinear frequency conversion protected by topology. Indeed, double-resonant corner modes have been exploited to significantly boost the second harmonic generation \cite{Chen21PRB_SHG,Om21PSS-PRL}, e.g., in \cite{Chen21PRB_SHG}, by matching two corner states within two different frequency bandgaps, efficiency
that is robust against defects and as high as 5.4$\times 10^{-3}$ W$^{-1}$ has been demonstrated (see Fig.\ref{figs:fig8}g). Furthermore, one could also couple the first-order topological edge states and the second-order topological corner states as demonstrated in \cite{Shi21OL_edgecorner} to realize more interesting applications, e.g., in \cite{Ma21AnnPhys_shg}, by making use of both the advantages of corner and edge states, i.e., the nonradiative characteristics of the corner states could be utilized to enhance the localized intensity for second harmonic generation whereas the topologically protected transmission of edge states could be exploited to transport the generated harmonic signal, the authors studied an interesting scenario in which the frequency of edge state is twice that of the corner state and demonstrated that harmonic wave generated from the fundamental corner mode could efficiently propagate along the system edge rather than spread into the bulk. Coupling between edge states \cite{Li22PhoRes_3layer} or corner states \cite{Zhang21JO_cornerarray} could lead to new phenomena, e.g., in \cite{Zhang21JO_cornerarray}, lattice topological edge and corner modes were proposed in a photonic crystal slab consisting of a periodic array of supercells (see Fig.\ref{figs:fig8}h), each of which hosts nontrivial edge and corner states. This may find possible applications in quantum-information processes, e.g.,  by placing quantum emitters into array of topological corner states, robust strong coupling and entanglement between these emitters could be fulfilled with the assistance of topological edge states \cite{Wang20PRapp_emitter}. The recently proposed dual-polarization topological corner states for both TE and TM modes \cite{Chen21PRapp_dualpol} and a new principle for creating corner states within odd-order bandgaps in $C_{4v}$-symmetric lattices  beyond the 2D SSH paradigm \cite{ChenNanoPhot22_oddgap} open new possibilities for both fundamental science and promising applications.

\begin{figure}
\includegraphics[width=\columnwidth]{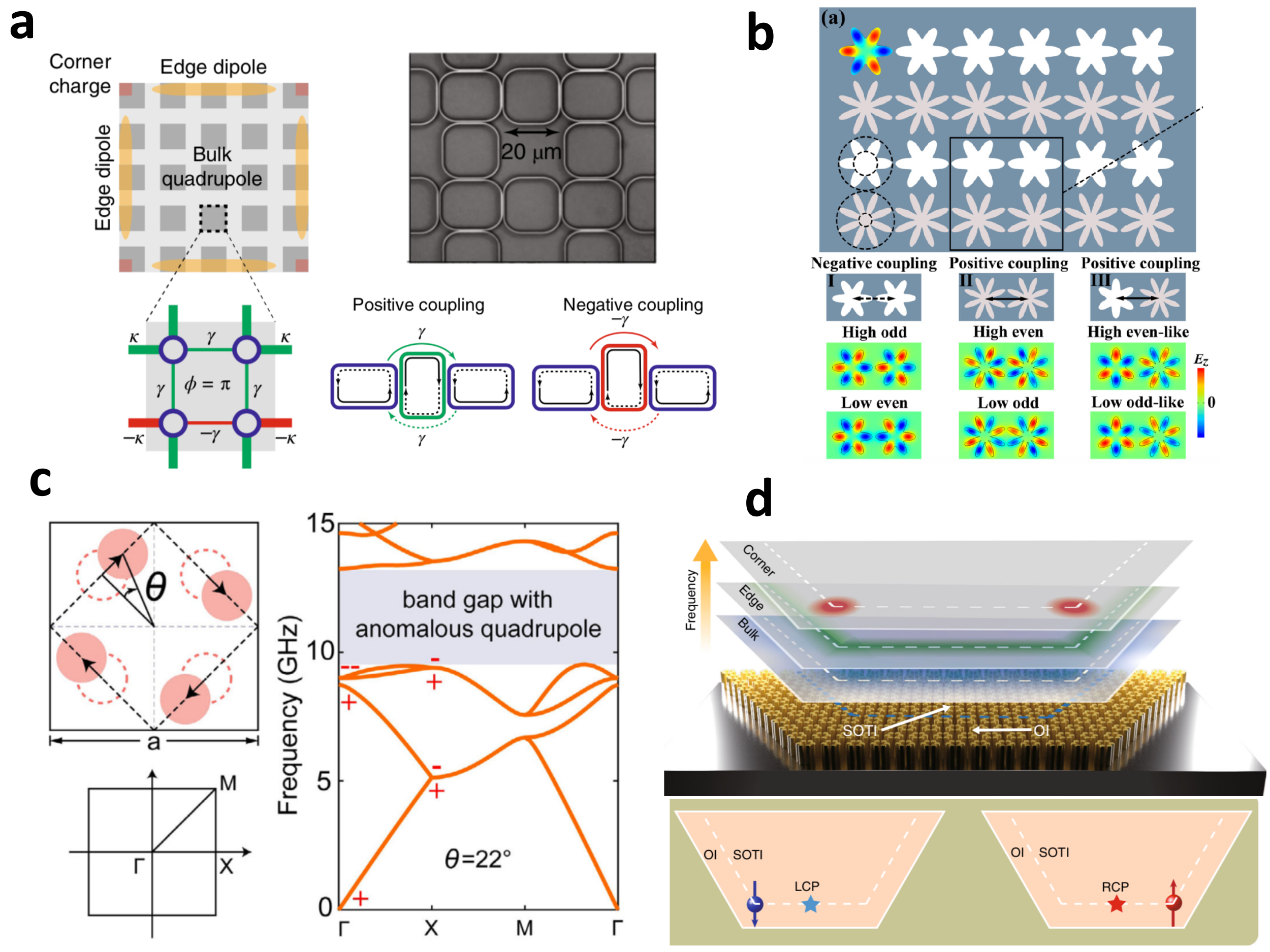}
\caption{Photonic quadrupole topological corner states. (a) Experimental implementation of the $\pi$-flux SSH model in a 2D lattice of nanophotonic silicon ring resonators. Reproduced with permission from~\cite{Mittal19NatPho_quadrupole}, Copyright (2019) by Springer Nature. (b) Theoretical proposal of the $\pi$-flux SSH model in a lattice of plasmon-polaritonic nanocavities. Reproduced with permission from~\cite{Chen20PRB_quadrupole}, Copyright (2020) by the American Physical Society. (c) Experimental observation of twisted quadrupole topological phases in photonic crystals made of dielectric cylinders via twisting the unit-cell. Reproduced with permission from~\cite{Zhou20LPR_twist}, Copyright (2020) by John Wiley and Sons. (d) Directional localization of photons at corners with opposite pseudospin polarizations. Reproduced with permission from~\cite{Xie20NC_Quadrupole}, Copyright (2020), under CC BY 4.0.  }
\label{figs:fig9}
\end{figure}

Before ending this section, we would like to briefly discuss 2D corner states stemming from nontrivial bulk quadrupole moment without dipole moment. This kind of insulators was called quadrupole insulators in the literature \cite{Benalcazar17Science} and it is challenging to implement the original $\pi$-flux SSH model proposed in \cite{Benalcazar17Science} using photonic systems due to the existence of negative couplings in the model.  Up to now, there are only a few works investigating quadrupole topological states in photonic systems \cite{Mittal19NatPho_quadrupole,Chen20PRB_quadrupole,Dutt20Light_synth,He20NC_Quadrupole, Liu19PRL_Quadrupole,Zhou20LPR_twist,Xie20NC_Quadrupole}. In \cite{Mittal19NatPho_quadrupole}, the authors experimentally implemented the negative coupling in a 2D lattice of nanophotonic silicon ring resonators (see Fig.\ref{figs:fig9}a) and demonstrated that the quantization of the bulk quadrupole moment manifests as topologically robust 0D corner states. Negative coupling could also be introduced in a lattice of plasmon-polaritonic nanocavities exploiting the geometry-dependent sign reversal of the couplings between the daisylike nanocavities \cite{Chen20PRB_quadrupole} (see Fig.\ref{figs:fig9}b) or in an array of modulated photonic cavities exploiting the idea of synthetic dimensions \cite{Dutt20Light_synth}. Quadrupole topological phases have also been theoretically studied in gyromagnetic photonic crystal through a double-band-inversion process \cite{He20NC_Quadrupole}. More interestingly, quadrupole topological phases could be realized in all-dielectric photonic crystals without the $\pi$-flux-threading mechanism. Denoting $P^n_i =q^n_i/2$ for the $n$th band, one could define a quadrupole through the dipole moments of all the bands below the bandgap as \cite{Liu19PRL_Quadrupole}

\begin{gather}
Q_{ij}=\sum_n^{N_{\text{occ}}}P^n_iP^n_j \hspace{0.2cm}  \text{mod} \hspace{0.2cm} 1
\end{gather}
which means that when the number of pairs of parities with opposite signs at $\Gamma$ and $X$ (or $M$ for hexagonal symmetry) of the same band below the bandgap is equal to $4m+2$ ($m=0,1,\cdots$), then the bandgap hosts a nontrivial quadrupole moment of $Q_{12}=1/2$ (note that the quadrupole is only well defined with vanishing dipole moment, which excludes the cases of $4m+1$ and $4m+3$). Quadrupole topological corner states have been experimentally observed in photonic crystals composed of dielectric cylinders via twisting the unit-cell \cite{Zhou20LPR_twist} as shown in Fig.\ref{figs:fig9}c or via the expanding/shrinking scheme \cite{Xie20NC_Quadrupole} as shown in Fig.\ref{figs:fig9}d.

%%%%%%%%%%%%%%%%%%%%%%%%%%%%%
\section{Topological photonics in 3D} \label{sec5}
%%%%%%%%%%%%%%%%%%%%%%%%%%%%%%

Photonic topological phases in 3D in general could be classified as gapless or gapped \cite{Xie21Frontier_3Drev}. For gapless topological phases, two or more energy bands are degenerate at certain points of the Brillouin zone, which may form points, lines or surfaces and so on. In the following, we briefly discuss gapless photonic topological phases related to Weyl points (twofold degeneracies), Dirac points (fourfold degeneracies), nodal lines and gapped 3D photonic topological insulators.

Weyl points are twofold degenerate points between two energy bands crossing linearly in momentum space. They behave as sources or sinks of the Berry flux in momentum space and must exist in negative/positive pairs in order to maintain the neutrality of the whole Brillouin zone. Since the Berry curvature is zero at every $\mathbf{k}$ for systems with PT symmetry, either P or T symmetry (or both) must be broken in order to obtain Weyl points, which are very different from 2D Dirac points, where either P or T symmetry breaking can easily gap out the 2D Dirac points.  Weyl points could be described by the Weyl Hamiltonian,
\begin{gather}
H_{\text{Weyl}}(\mathbf{k})=v_xk_x\sigma_x + v_yk_y\sigma_y +v_zk_z\sigma_z
\end{gather}
where $v_i$ is the group velocity and $\sigma_i$ the Pauli matrix. The topological invariant of Weyl point could be calculated by the integral of the Berry curvature over a surface enclosing the Weyl point, $C_{\text{Weyl}}=\frac{1}{2\pi}\int_S F_{\text{Weyl}}(\mathbf{k})dS$ and a simple calculation gives $C_{\text{Weyl}}=\text{sgn}(v_xv_yv_z)$. It is to be noted that 3D Weyl points are absolutely robust against any perturbations, which could be understood from the fact that all three Pauli matrices are used in the Hamiltonian and thus no term can open a gap for $H_{\text{Weyl}}$. Consequently, the only way to annihilate a Weyl point is when two oppositely charged Weyl points meet in momentum space.

\begin{figure*}
\includegraphics[width=1\textwidth]{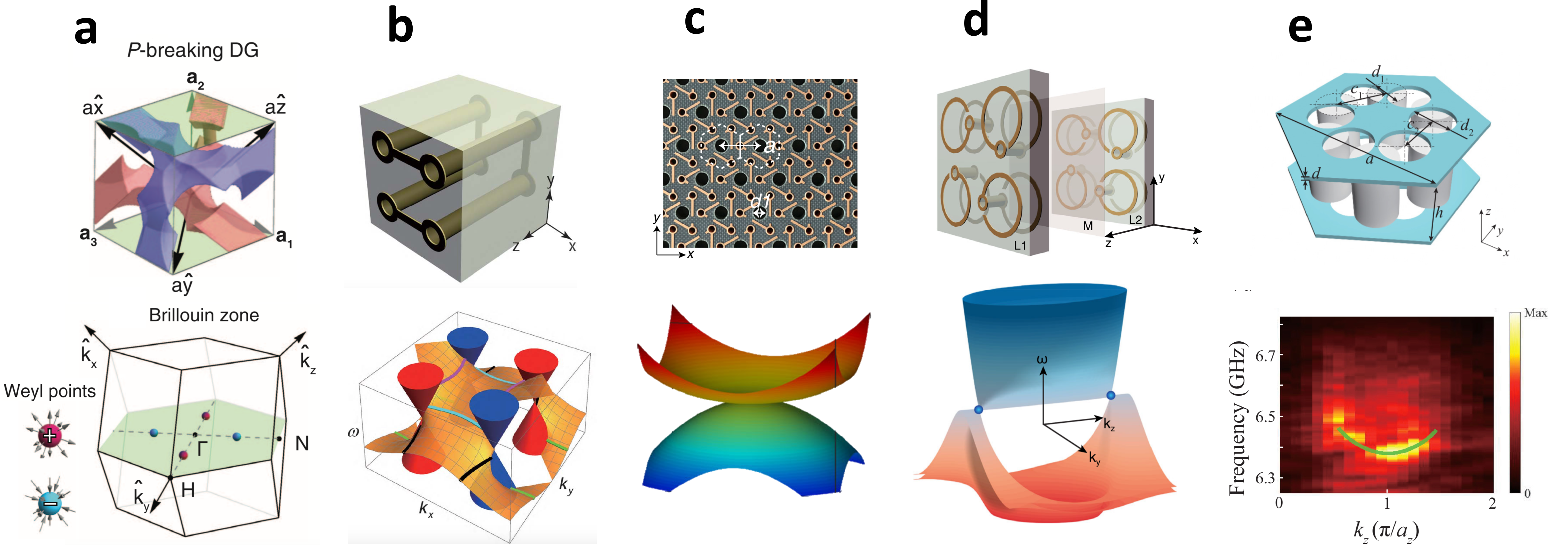}
\caption{Various nodal points in 3D photonic systems. (a) Experimental observation of Weyl points in a double-gyroid photonic crystal with inversion-breaking. Reproduced with permission from~\cite{Lu15Science_weyl}, Copyright (2015) by the American Association for the Advancement of Science. (b) Ideal Weyl points and helicoid surface states in a microwave photonic crystal of saddle-shaped metallic coils. Reproduced with permission from~\cite{Yang18Science_ideal}, Copyright (2018) by the American Association for the Advancement of Science. (c) Ideal unconventional charge 2 Weyl point in a chiral microwave metamaterial made from printed circuit boards. Reproduced with permission from~\cite{Yang20PRL_unconven}, Copyright (2020) by the American Physical Society. (d) Photonic Dirac points in an elaborately designed metamaterial. Reproduced with permission from~\cite{Guo19PRL_surface}, Copyright (2019) by the American Physical Society. (e) Higher-order Dirac semimetal and the hinge state dispersion in coupled layers of deformed photonic honeycomb lattices. Reproduced with permission from~\cite{Wang22PRB_high_order}, Copyright (2022) by the American Physical Society.}
\label{figs:fig10}
\end{figure*}

Weyl points of photons were first studied in double-gyroid photonic crystals \cite{Lu13NP_pointLine}. Starting from a threefold degeneracy at the Brillouin zone centre, the authors showed that Weyl points could be obtained by perturbations breaking either P or T. It is to be noted that the minimal number of Weyl points in systems breaking T symmetry is two, while in systems respecting T symmetry it is four, because T maps a Weyl point at $\mathbf{k}$ to $-\mathbf{k}$ with the same chirality whereas P maps a Weyl point at $\mathbf{k}$ to $-\mathbf{k}$ with the opposite chirality. Thus in systems respecting T, there must exist at least two other Weyl points of opposite chirality, to neutralize the whole system. Weyl points have been theoretically studied in magnetized plasma \cite{Gao16NC_palsma}, magnetic photonic crystal \cite{Yang17OE_magnetic}, gyromagnetic metamaterials \cite{Li21PRB_gyromagnetic} and experimentally realized in magnetized semiconductor \cite{Wang19NatPhy_magnetSemicond}. As magnetic response in general is not easy or convenient to be implemented in photonic systems, breaking P to create Weyl points is an alternative. In fact, the first observation of Weyl points in photonics is via inversion-breaking in a double-gyroid photonic crystal \cite{Lu15Science_weyl} (see Fig.\ref{figs:fig10}a), where excited bulk states probed by angle-resolved microwave transmission measurements show two linear dispersion bands touching at four isolated points in the 3D Brillouin zone, indicating the observation of Weyl points. Later on, Weyl points in systems by breaking P but preserving T have also been theoretically proposed in photonic crystal superlattices \cite{Abad152Dmat} and chiral metamaterials \cite{Gao15PRL_chiralHyper, Xiao16PRL_hyperWeyl}. However, as the Weyl points in \cite{Lu15Science_weyl} are not symmetry-related, they occur at different frequencies. Symmetry-related Weyl points at the same frequency have been studied in modified double gyroid with $D_{2d}$ symmetry \cite{Wang16PRA_ideal} and experimentally demonstrated in a microwave photonic crystal of saddle-shaped metallic coils \cite{Yang18Science_ideal} (see Fig.\ref{figs:fig10}b), where these Weyl points were called ideal, not only because they all exist at the same frequency but also that they are separated from any other bands. Recently, Weyl points at optical frequencies have been experimentally observed in a bio-inspired 3D photonic crystal coated uniquely with layered-composite nanometric materials \cite{Goi18LPR_optical} and theoretically proved feasible via an interference lithographic design \cite{Park20ACSPho_visible}. Due to the bulk-edge correspondence principle, the bulk Weyl points can give rise to nontrivial surface state --- so-called Fermi arc, which connects the projections of Weyl points with non-vanishing charges on the surface Brillouin zone. The first experimental measurement of robust surface states was conducted in a microwave Weyl photonic crystal \cite{Chen16NC_weyl_arc} and later also in a chiral hyperbolic metamaterial \cite{Yang17NC_arc} as well as in laser-written waveguides at optical frequencies \cite{Noh17NatPhy_weyl_arc}. In contrast to the Weyl point with a pointlike Fermi surface and a vanishing density of states, Weyl point can also be associated with a strongly tilted cone dispersion that breaks the Lorentz invariance, i.e., the so-called type-II Weyl point, in which the Fermi surface consists of touched electron-hole pockets with nonvanishing density of states \cite{Soluyanov15Nature_type2weyl}. Type-II Weyl points have been demonstrated in different photonic systems, such as, photonic crystals \cite{Chen16NC_weyl_arc,Chen21OE_1Dtwist}, waveguide arrays \cite{Noh17NatPhy_weyl_arc,Qin18OE_type2weyl} and chiral metamaterials \cite{Xiao16PRL_hyperWeyl,Yang17NC_arc,Li22PRB_type2weyl}.

\begin{figure*}
\includegraphics[width=0.9\textwidth]{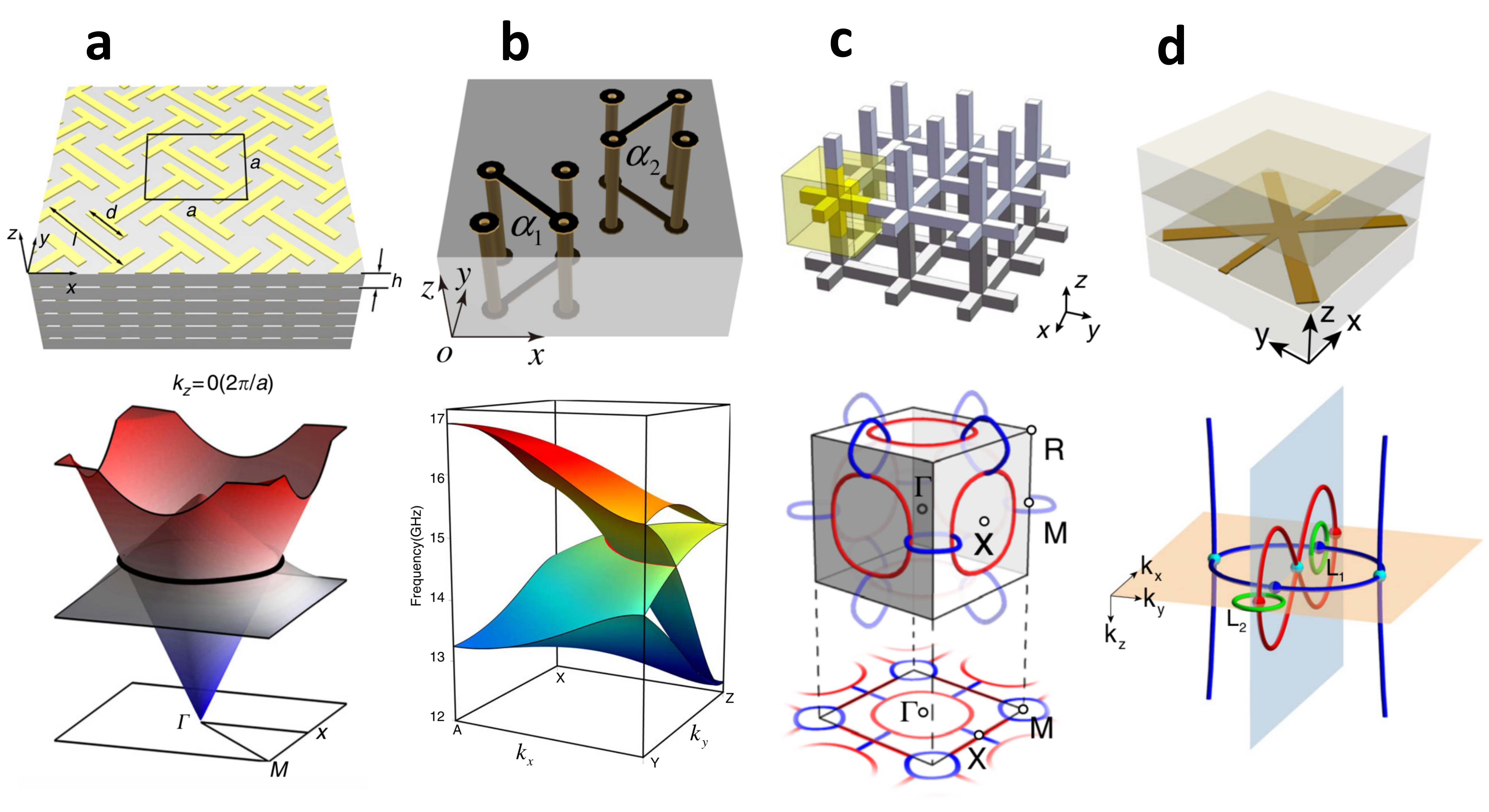}
\caption{Different kinds of 3D photonic nodal lines. (a) Photonic nodal lines in metacrystals. Reproduced with permission from~\cite{Gao18NC_line}, Copyright (2018), under CC BY 4.0. (b) Hourglass nodal lines in photonic metacrystals. Reproduced with permission from~\cite{Xia19PRL_hourglass}, Copyright (2019) by the American Physical Society. (c) Experimental observation of nodal chains in a metallic-mesh photonic crystal. Reproduced with permission from~\cite{YanNatPhy_chain}, Copyright (2018) by Springer Nature. (d) Non-Abelian nodal links in a biaxial hyperbolic metamaterial. Reproduced with permission from~\cite{Yang20PRL_link}, Copyright (2020) by the American Physical Society.}
\label{figs:fig11}
\end{figure*}

Apart from the Weyl points carrying a charge of $\pm 1$, Weyl points could also carry an arbitrary integer charge n, described by the following 3D Hamiltonian \cite{Fang12PRL_multiWeyl},
\begin{gather}
H_n(\mathbf{k})=k_+^{n}\sigma_+ + k_-^{n}\sigma_- + k_z\sigma_z + \omega_0 I
\end{gather}
where $k_{\pm}=(k_x\pm ik_y)$ and $\sigma_{\pm}=(\sigma_x\pm i\sigma_y)/2$, $I$ is the identity matrix and $\omega_0$ the frequency of the Weyl point. Charge 2 Weyl points have been proposed in woodpile photonic crystals \cite{Chang17PRB_woodpile,Takahashi18JPSJ_charge2} and been observed experimentally in the mid-infrared regime in a low-index chiral woodpile photonic crystal fabricated by two-photon polymerization \cite{Vaidya20PRL_charge2}. The splitting of the charge 2 Weyl point into two charge 1 Weyl points was further experimentally observed in \cite{Jorg22LPR_charge2} under careful symmetry breaking.  However, due to the low index contrast in \cite{Vaidya20PRL_charge2}, there is only an incomplete bandgap surrounding the Weyl point, making it challenging to study the Weyl point in isolation. An ideal charge 2 Weyl point separated from trivial bands was experimentally observed in \cite{Yang20PRL_unconven} (see Fig.\ref{figs:fig10}c), where two long surface arcs that form a noncontractible loop wrapping around the surface Brillouin zone were mapped out in the experiment. Very recently, a charge-4 Weyl point was demonstrated experimentally in \cite{Chen22arxiv_maxCharge}, where the authors mapped out the projected bulk dispersion and the exotic quadruple-helicoid Fermi arcs of the topological surface states emanating from the charge 4 Weyl points.

Engineering photonic systems that can exhibit Weyl points usually is complex and methods, such as, group theory \cite{Saba17PRL_group} or practical approach based on self-assembly \cite{Frucharta18PNAS_assembly} have been proposed. One can further exploit the idea of synthetic dimensions to explore Weyl physics in lower spatial dimensions  \cite{Lin16NC_synWeyl,Wang17PRX_synWeyl,Chen21OE_1Dtwist,Lee22PRL_2layer,Ma21Science_5D}, providing the potential for on-chip integration. For example, in \cite{Lin16NC_synWeyl}, the authors proposed to use 2D arrays of resonators undergoing dynamic modulation of refractive index to explore 3D Weyl physics, in which the non-trivial topology of the Weyl point manifests in terms of surface state arcs in the synthetic space exhibiting one-way frequency conversion. In \cite{Ma21Science_5D}, by using bianisotropic terms as the synthetic fourth and fifth dimensions, intriguing bulk and surface phenomena, such as linking of Weyl surfaces and surface Weyl arcs in 5D have been studied. Weyl point physics could also find interesting applications, e.g., in \cite{Jia19Science_zero}, the chiral zero mode of Weyl points, which is a one-way propagating bulk mode, has been applied for the robust transport of photons in the bulk medium. In \cite{Cheng20PRL_spiral}, generation of vortex beam has been demonstrated using the reflection property of Weyl metamaterials.

A Dirac point in 3D is a fourfold degenerate point, which could also be viewed as two $\pm 1$ charged Weyl points  located at the same position in momentum space. 3D photonic Dirac points have been studied in photonic crystals \cite{Wang16PRB_pointGroup} and metamaterials \cite{Guo17PRL_dirac}. In \cite{Wang16PRB_pointGroup}, the authors found a pair of stable 3D Dirac points in hollow cylinder hexagonal photonic crystals and demonstrated that $C_6$ is the only point group symmetry that can stabilize the paired Dirac points in 3D photonic crystals. In \cite{Guo17PRL_dirac}, the authors theoretically showed that Dirac points can also be realized in effective media under electromagnetic duality and found that a pair of spin-polarized Fermi-arc-like surface states emerges at the interface between air and the Dirac metamaterials. Furthermore, type-II Dirac photons where the linear crossings forming the degenerate points are tilted, were shown could exist in photonic crystal with nonsymmorphic screw symmetry \cite{Wang17npjQM_type2}. 3D photonic Dirac points and their spin-polarized surface arcs have been experimentally observed in the microwave region with an elaborately designed metamaterial \cite{Guo19PRL_surface} (see Fig.\ref{figs:fig10}d), where two symmetrically placed Dirac points are stabilized by electromagnetic duality symmetry. As a Dirac point consists of two Weyl points with opposite charges, two Fermi arcs exhibiting left (right) circular polarization with the plane of the polarization being parallel to the Fermi arcs could emerge from the Dirac point. Beyond this conventional bulk-edge correspondence, higher-order hinge states connecting the momentum-space projections of the two Dirac points localized at the hinge have been measured experimentally at microwave frequencies \cite{Wang22PRB_high_order} (see Fig.\ref{figs:fig10}e). Recently,  a charge-2 Dirac point was experimentally demonstrated by deliberately engineering hybrid topological states in a 1D optical superlattice system using the idea of synthetic dimensions  \cite{Hu20ComPhy_charge2}. When a gaussian beam is reflected near photonic Dirac point at an optical interface, vortical phase distribution and spin inversion could happen \cite{Xu21PRA_poincare}, providing an interesting way to manipulate polarized vortex beam with photonic Dirac point.

We would like to note that Fermi arcs in Weyl semimetals are topologically protected and thus are robust because of the nonzero and opposite chirality of the two Weyl points that the Fermi arcs connect. On the other hand, as a Dirac point consists of two degenerate Weyl points with opposite chirality, there exist two branches of Fermi arcs (double Fermi arcs) connecting a pair of 3D Dirac points. However, the double Fermi arcs connecting the two Dirac points are not topologically protected due to the zero chirality of the Dirac points and thus can be continuously deformed into a closed Fermi contour without any symmetry breaking \cite{Kargariana16PNAS}. Moreover, Weyl points carrying higher topological charges can have multiple Fermi arcs emanating from them, which can stretch over a large portion of the Brillouin zone or even form a noncontractible loop winding around the surface Brillouin zone \cite{Yang20PRL_unconven}, unlike the single, short and open Fermi arc connecting two charge $\pm$1 Weyl points.

 \begin{figure*}
\includegraphics[width=1\textwidth]{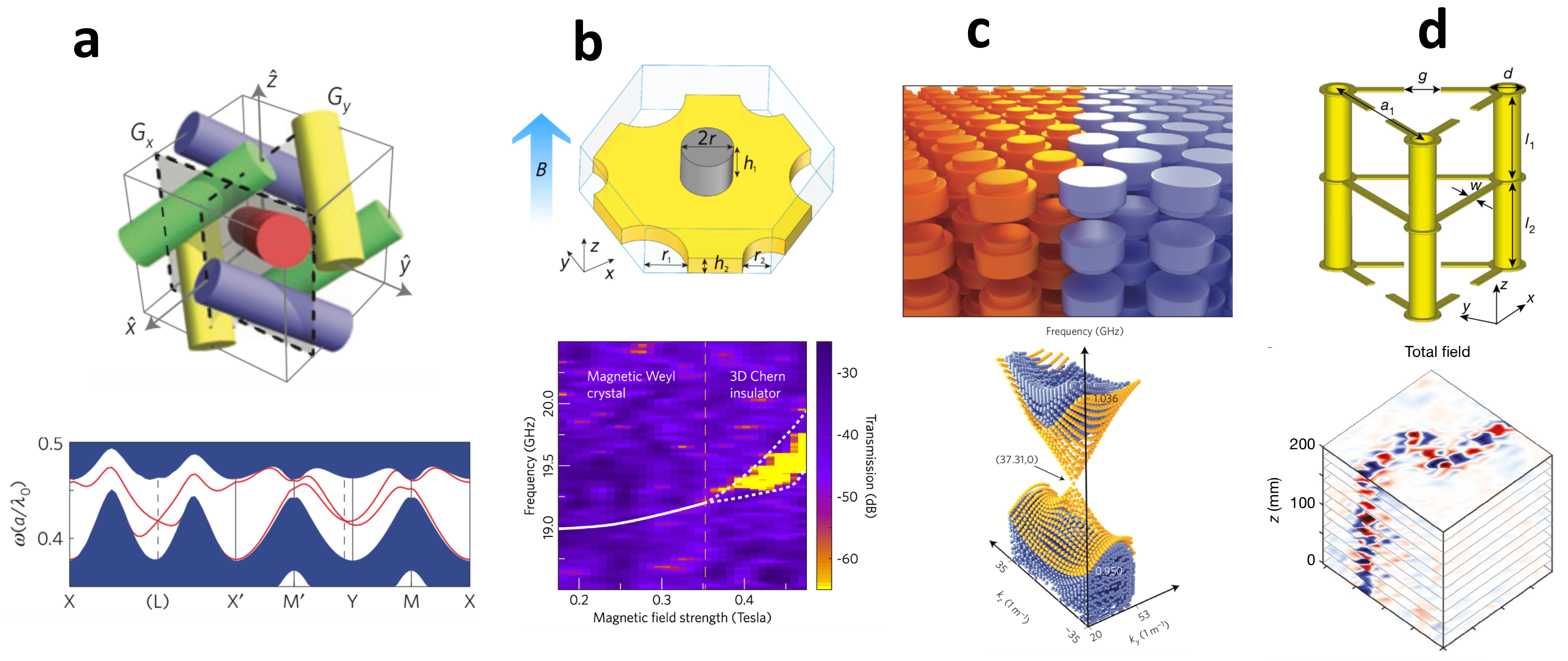}
\caption{Gapped 3D photonic topological insulators. (a) Topological photonic crystal hosting a single surface Dirac cone at L protected by the nonsymmorphic glide reflection. Reproduced with permission from~\cite{Lu17NP_3d}, Copyright (2016) by Springer Nature. (b) Weyl point pair annihilation and  bandgap opening in a gyromagnetic photonic crystal~\cite{Liu21arxiv_weylAnn}. (c) Domain wall in an all-dielectric bianisotropic metacrystal and the Dirac-like dispersion of the surface states. Reproduced with permission from~\cite{Slobozhanyuk17NP_3d}, Copyright (2017) by Springer Nature. (d) Photonic topological insulator with robust photonic propagation along a non-planar surface in a photonic structure made from split-ring resonators. Reproduced with permission from~\cite{Yang19Nature_3d}, Copyright (2019) by Springer Nature. }
\label{figs:fig12}
\end{figure*}

In addition to the point degeneracies discussed above, the band crossings could also form 1D line, so-called nodal line. Depending on the shape of the nodal line, different configurations \cite{Park22NanoPhot_3Drev}, such as, nodal ring, nodal chain, nodal link and nodal knot, could exist. The surface states corresponding to these 1D bulk  topological states are much richer than the Fermi arc states associated with Weyl/Dirac points. Photonic nodal line in the form of a single ring was experimentally demonstrated in microwave cut-wire metacrystals \cite{Gao18NC_line} (see Fig.\ref{figs:fig11}a), where  both the toroidal bulk state and the drumhead surface state supported by the metacrystal were verified. Later on, a glide mirror symmetry protected hourglass nodal line formed by an hourglass-shaped band dispersion around a loop in the momentum space was experimentally demonstrated in a photonic metacrystal at microwave frequency \cite{Xia19PRL_hourglass} (see Fig.\ref{figs:fig11}b), where the observed hourglass nodal line resides in a clean and large frequency interval and is immune to symmetry preserving perturbation. Ideal nodal ring has also been experimentally observed in a simple 1D photonic crystal in visible region \cite{Deng21arxiv_ring}. Dirac nodal line semimetal with fourfold line degeneracy and its perpendicularly polarized double-bowl surface states were further experimentally observed in \cite{Hu21Light_2bowl}.  Nodal lines could also act as quadrupole sources of Berry curvature flux \cite{Wang22PRL_QBC}, or intersect at hidden-symmetry-enforced nexus points \cite{Xiong20light_nexus}. Apart from a single nodal ring, two or more rings could chain together via touching points forming nodal chains. Nodal chains have been experimentally studied in metallic-mesh photonic crystals \cite{YanNatPhy_chain,Wang21SciRep_drumhead} (see Fig.\ref{figs:fig11}c) and bi-anisotropic metamaterials \cite{Wang21light_quaternion}. When the nodal line is formed with more than 2 band degeneracies,  the nodal lines formed by consecutive pairs of bands exhibit interesting braiding structures and the underlying topological charges could be described by quaternions. Non-Abelian nodal links formed by the crossings between three adjacent bands were experimentally demonstrated in a biaxial hyperbolic metamaterial \cite{Yang20PRL_link} (see Fig.\ref{figs:fig12}d). Recently, a dielectric photonic crystal in the form of double diamond structure was theoretically shown to host a nodal link with non-Abelian charges \cite{Park21ACSpho_non-abelian}. Non-Abelian frame charges can flow in momentum space along nodal lines, which has been observed in experiments using a biaxial photonic crystal \cite{Wang22arxiv_frame}. Band crossings can also form 2D nodal surface, e.g., as shown in \cite{Kim19PRB_surface}.

Finally, we would like to briefly discuss the gapped photonic topological phases. In general, 3D photonic Chern insulators could be characterized by three first Chern numbers, i.e., a Chern vector $\mathbf{C}=(C_x,C_y,C_z)$, defined on lower dimensional surfaces \cite{Oono16PRB_sectionChern,Devescovi21NC_chern}. The above discussed gapless degeneracies, such as Weyl/Dirac points, or nodal lines could be gapped to generate a nontrivial bandgap by changing the parameters of the systems, e.g., in magnetic photonic crystals, nontrivial bandgap could be obtained by gapping out 3D Dirac points as theoretically studied in \cite{Lu17NP_3d,Kim21OE_glide} (Fig.\ref{figs:fig12}a) and very recently, in a microwave-scale gyromagnetic 3D photonic crystal, the authors in \cite{Liu21arxiv_weylAnn} found that momentum space locations of a single pair of ideal Weyl points strongly depend on the biasing magnetic field and by continuously varying the field strength, annihilation of the Weyl points and thus formation of a 3D Chern insulator were observed in the experiments (see Fig.\ref{figs:fig12}b). Gapped photonic topological phases have also been studied in all-dielectric bianisotropic metacrystal \cite{Slobozhanyuk17NP_3d} (Fig.\ref{figs:fig12}c). The first fully 3D topological photonic bandgap was experimentally realized in 2019 based on a 3D array of metallic split-ring resonators at microwave frequencies \cite{Yang19Nature_3d} (see Fig.\ref{figs:fig12}d), which was achieved by gapping out 3D Dirac points via the bi-anisotropic response of the split-ring resonators. Moreover, by direct field measurements, both the gapped bulk band structure and the Dirac-like dispersion of the photonic surface states were mapped out successfully in the experiment.

%%%%%%%%%%%%%%%%%%%%%%%%%%%
\section{Conclusions and discussions}\label{sec6}
%%%%%%%%%%%%%%%%%%%%%%%%%%%

In summary, we have reviewed the recent developments of topological photonics in one, two and three dimensions. In specific, in 1D we presented the paradigmatic SSH model for topological physics, illustrated its topological features, such as the winding number, through an intuitive tight-binding model and discussed various photonic platforms that have been used to implement the SSH model for different topology-related applications. In 2D, we discussed four categories of photonic topological states, i.e., the quantum Hall states, quantum spin Hall states, quantum valley Hall states and second-order topological corner states, where the topological invariants for these four different states are Chern number, spin Chern number, valley Chern number and bulk dipole or quadrupole moment, respectively. In 3D, the photonic topological phases in general could be classified as gapped or gapless. For gapped photonic systems, if the T symmetry is broken, one could have 3D photonic Chern insulators characterized by a Chern vector. For gapless photonic topological phases, if the band crossings form isolated points in the Brillouin zone, one could have nodal points, such as twofold degenerate Weyl points, or fourfold degenerate Dirac points, whose topological invariants are the topological charges carried by the degenerate points and one could also have type-I  or type-II nodal points depending on the slope of the bands near the crossing. The band crossings could also form 1D degenerate nodal lines, and depending on the shape and number of the nodal lines, one could have nodal ring, nodal chain, nodal link, and nodal knot,  where the Berry phase along a close loop enclosing the line could serve as the topological invariant for nodal lines. Moreover, if the nodal lines are from band crossings of more than two bands, one could have nontrivial non-Abelian braiding behaviors, hosting non-Abelian topological charges.

Regarding future perspectives, we believe the 1D SSH model will continue attracting attention due to its simplicity and transparent physics. One could extend the model, such as to consider trimer or tetramer SSH models and even to include longer-range hopping, to enrich the physics this model can host. Nonetheless, as the topological states are 0D in this model, they can not be used to transport electromagnetic waves. In 2D, the most robust photonic topological states are quantum Hall states with time-reversal symmetry breaking \cite{Haldane08PRL,Wang09nature}. In this case, as the backscattering channels are completely removed, the one-way edge modes are absolutely robust as long as the disorder is not strong enough to close the bandgap. However, in order to break time-reversal symmetry, one would need real or synthetic magnetic fields. While real magnetic fields could be used in microwave regimes, magnetic response in general is weak in visible light regimes, where synthetic magnetic fields would be preferred \cite{Fang12NatPho_magnetic,Hafezi11NatPhy_resonator}. To emulate quantum spin Hall states for light, one could consider different combinations of TE and TM polarizations. To do this, in general, one would need to first create a double Dirac one and then find ways to gap out the double Dirac cone. One point to be noted is that the robustness of the resulting quantum spin Hall states is different depending on the specific mechanism to implement the double Dirac cone and the way to open the bandgap, e.g., the quantum spin Hall states based on bi-anisotropic response to open the bandgap in general is more robust than the method based on lattice symmetry consideration \cite{Khanikaev13NatMat,WuHu15PRL}. While quantum spin Hall states need a double Dirac cone, a single Dirac cone could be exploited to emulate quantum valley Hall states. It would be interesting to find a way such that the valleys could be tuned to any desired locations in the Brillouin zone rather than pinned to the $K/K'$ points due to lattice symmetry. Apart from these conventional topological states, second-order topological phases with corner states have also attracted a great attention in recent years. Up to now, most of the photonic corner states studied in the literature are based on nontrivial bulk dipole moments. For corner states based on nontrivial bulk quadrupole moments, the original $\pi$-flux square lattice SSH model \cite{Benalcazar17Science} is not very convenient for photonic systems, because one would need to implement the negative coupling in the model. So finding new lattice structures and tight binding models without negative coupling that could be implemented using all-dielectric materials is an interesting direction. It would also be interesting to study the difference between the corner states created using bulk dipole and quadrupole moments, such as, do they have the same degree of robustness or one kind of corner states has a stronger robustness than the other? In 3D, the topological physics is very rich \cite{Vergniory22Sci_3Dtopo}, and most of the developments in topological photonics are motivated by the relevant topological phenomena in condense matter physics. Up to now, most studies in 3D topological photonics have focused on conventional bulk-edge correspondence and Weyl nodes as well as their surface states have already found many interesting applications, e.g., one-wave waveguide \cite{Jia19Science_zero,Kim19AOM_broadband}, frequency selectivity \cite{Wang16PRA_ideal}, vortex beam generation \cite{Cheng20PRL_spiral}, topological self-collimation \cite{Yang20PRL_unconven},  optical tweezers \cite{Yang22NJP_tweezer}, cloaking \cite{Takahashi21OE_cloaking}, superimaging \cite{Wang22PRL_QBC}, all-angle negative refraction and Veselago imaging \cite{Yang21optica_Veselago}. Higher-order topological photonic states in 3D, such as third-order corner states or second-order hinge states, are less explored and many topological phenomena difficult to implement in electronic materials are waiting to be studied in the photonics context. The possible problem is that as 3D photonic systems are bulky, they probably will not be compatible to the future practical on-chip applications. One possible solution is to exploit lower dimensional photonic systems, such as 1D or 2D, to emulate the 3D (or even higher dimensions) topological physics using the idea of synthetic dimensions \cite{Yuan18optica_synD,Lustig21AOP_synD}. Then a possible complication in this case would be what is the consequence of the topological phenomena in real spatial dimensions rather than the synthetic dimensions, as in reality, devices are always operated in real space. We believe topological photonics will continue attracting the interests of photonics researchers in the years to come due to its interesting physics and promising practical applications.

\section{Acknowledgements}
This work was supported in part by the National Natural Science Foundation of China (Nos. U20A20164 and 61975177).

\end{document}